\documentclass{article}

\usepackage[utf8]{inputenc}
\usepackage[T1]{fontenc}
\usepackage{lmodern}

\usepackage[english]{babel}
\usepackage{csquotes}

\usepackage[a4paper, margin=1in]{geometry}

\usepackage{graphicx}
\usepackage{subcaption}
\usepackage{float}
\usepackage{amsmath}
\usepackage{amssymb}

\usepackage[ruled,vlined]{algorithm2e}

\usepackage[table]{xcolor}
\usepackage{tabularx}
\usepackage{longtable}
\usepackage{array}
\usepackage{booktabs}
\usepackage{multirow}
\usepackage{rotating}
\usepackage{ragged2e}

\definecolor{myyellow}{RGB}{255,255,102}
\definecolor{myorange}{RGB}{226,136,70}
\definecolor{myred}{RGB}{235,51,35}
\definecolor{mygreen}{RGB}{165,210,100}

\usepackage[
  style=apa,
  backend=biber,
  language=english
]{biblatex}

\DeclareLanguageMapping{english}{english-apa}
\usepackage{hyperref}
\hypersetup{
  colorlinks=false,
  pdfborderstyle={/S/U/W 1.5},
  linkbordercolor=blue,
  citebordercolor=blue,
  urlbordercolor=blue
}

\usepackage[nameinlink]{cleveref}

\crefname{figure}{Figure}{Figures}
\Crefname{figure}{Figure}{Figures}
\crefname{table}{table}{tables}
\Crefname{table}{Table}{Tables}

\usepackage{titlesec}
\titleformat{\section}{\large\bfseries}{\thesection.}{0.5em}{}
\titleformat{\subsection}{\normalsize\bfseries}{\thesubsection.}{0.5em}{}

\usepackage{authblk}

\title{Polycentrality in Urban Socioecological Systems: A Multilayer Analysis of Social Infrastructure in Iztapalapa, Mexico City}

\author{Marco A. Rosas\textsuperscript{1}}
\author{Gustavo Carreón\textsuperscript{2}}
\author{Carlos Gershenson\textsuperscript{3}}
\author{Jesús Mario Siqueiros\textsuperscript{4}}

\affil{\small\textsuperscript{1} Center for Complexity Sciences (C3), UNAM}
\affil{\small\textsuperscript{2} Institute of Economic Research (IIE), UNAM}
\affil{\small\textsuperscript{3} Binghamton University, SUNY}
\affil{\small\textsuperscript{4} Institute for Research in Applied Mathematics and Systems (IIMAS), UNAM}

\newcounter{backref}

\date{}

\begin{document}

\maketitle

\begin{abstract}

The spatial distribution of infrastructure and activities, along with the networks and interactions that connect them, is key to understanding the relationship between urban form, function, and sustainability from a complex systems perspective. Within this dynamic relationship, the extent to which localized social infrastructure can reinforce existing functional centralities or create conditions for the emergence of new ones remains insufficiently understood.

This study examines these relationships through the UTOPÍAS program in Iztapalapa, Mexico City, where 15 public social facilities have been established across a historically underserved urban territory. We develop a multilayer network framework integrating pedestrian infrastructure, GTFS-based public transport services, and UTOPÍAS as a target layer to assess multimodal connectivity, travel-time accessibility, and spatial coverage. These conditions are examined alongside indicators of local economic centrality (PCI and wPCI) to investigate how infrastructure interdependencies relate to the functional organization of urban space.

The analysis reveals spatial heterogeneity in multimodal accessibility and differences in the integration of UTOPÍAS with existing transport networks. By relating these patterns to pre-existing economic centralities, we examine whether distributed social infrastructure may reinforce established functional centers or create conditions favorable to emerging centralities, without assuming that spatial decentralization necessarily produces polycentrality.

The findings contribute a relational, data-driven approach to understanding how social infrastructure and mobility shape urban functions, emphasizing accessibility and territorial integration as critical conditions for more equitable and sustainable urban development.

\end{abstract}


\section{Introduction}

Cities are complex urban socioecological systems in which social, infrastructural, environmental, and spatial processes interact across scales \textcolor{blue}{\parencite{Batty2005, Bettencourt2013, Frank2017, Cioffi2016}}. These interactions shape heterogeneous urban configurations and the conditions under which activities and functions develop. Urban form is therefore more than the physical arrangement of the built environment: it organizes opportunities, mobility requirements, and relationships among populations and infrastructures. Its relevance to sustainability lies in how spatial configurations enable or constrain these relationships.

The distribution of activities influences travel requirements, accessibility, and the social and environmental costs of mobility \textcolor{blue}{\parencite{Banister2005, Sonnberger2021, Urry2009}}. Dispersed or weakly integrated urban structures may reinforce long journeys and unequal access, whereas better-integrated configurations can support proximity and interaction \textcolor{blue}{\parencite{BautistaHernandez2020}}. Polycentrality has consequently been discussed as an alternative to strongly monocentric organization, with the potential to redistribute opportunities among multiple centers \textcolor{blue}{\parencite{Chen2021, Finka2015}}. Yet morphological approaches based on population, employment, density, and economic concentrations \textcolor{blue}{\parencite{Xu2023}} do not establish whether these locations function as centers within the wider urban system.

Functional polycentricity instead concerns the interactions connecting centers, populations, and activities. Flow- and accessibility-based approaches complement morphological descriptions by examining how urban centers attract and organize interactions \textcolor{blue}{\parencite{Veneri2010, Wu2021}}. This distinction matters for urban change: new activities may reinforce established centralities or develop in structurally less central locations, without necessarily transforming metropolitan morphology. The question is therefore not simply where facilities are distributed, but how their relationships with existing urban structures can support localized functional centralities.

Social infrastructure provides a relevant setting for examining this question. Public facilities and institutions create spaces for social interaction, collective life, and access to shared services \textcolor{blue}{\parencite{Hall2020, Zahnow2024}}, although their functional significance depends on the populations and interactions they support \textcolor{blue}{\parencite{Högström2022, Kim2024}}. Spatial decentralization can introduce opportunities outside established centers, but does not by itself redistribute urban function. Whether a facility reinforces an existing centrality or contributes to a potentially emerging one depends on its integration with surrounding populations, activities, and mobility networks.

Mobility and accessibility mediate that integration. Transport systems, land use, and activity locations must be examined together \textcolor{blue}{\parencite{Banister2005}}: pedestrian and public transport networks shape opportunities to reach social infrastructure, while observed use indicates whether some of these opportunities become realized interactions. Accessibility and visitation are thus complementary, but neither independently establishes functional centrality. Their relationship with pre-existing urban structure provides a basis for examining heterogeneous form--function configurations.

Iztapalapa, a densely populated borough of Mexico City \textcolor{blue}{\parencite{INEGI2017}}, offers a pertinent case because longstanding inequalities in access to urban opportunities coexist with investments in social infrastructure and multimodal transport. Since 2019, the UTOPÍAS program (Units for Transformation and Organization for Inclusion and Social Harmony) has introduced publicly funded facilities offering free cultural, educational, sports, environmental, and community services across historically underserved neighborhoods. This study examines 15 UTOPÍAs distributed across different structural and accessibility contexts, within a metropolitan setting characterized by a predominantly monocentric urban organization.

We examine the coupling of pre-existing urban centrality, social infrastructure, multimodal accessibility, and realized use to distinguish configurations consistent with reinforced, emerging, or functionally constrained centralities. Their collective spatial organization also raises the question of whether localized functional differences are consistent with patterns of functional polycentricity at the local scale. Accordingly, this study asks: \textbf{How do pre-existing urban structure, social infrastructure, and multimodal accessibility relate to the reinforcement or potential emergence of functional centralities, and what do these relationships reveal about localized polycentricity and its implications for urban sustainability in Iztapalapa?}

To address this question, we characterize pre-existing centrality and construct a multilayer representation connecting pedestrian mobility, public transport, and UTOPÍAs. We examine multimodal travel times, connectivity, and population coverage, and compare these potential accessibility conditions with independent observations of visitation and territorial reach. By keeping structural position, accessibility, and realized function analytically distinct, the study evaluates their convergence and divergence rather than assuming that infrastructure provision generates new centers.

The contribution is a relational framework for assessing how social infrastructure becomes functionally embedded in heterogeneous urban settings. Polycentricity is not assumed to be inherently sustainable: its potential benefits depend on the extent to which functional centralities support proximity, provide access to distributed opportunities, and offer alternatives to dependence on dominant centers across socially differentiated territories \textcolor{blue}{\parencite{Ostrom2001, Veneri2010, Yang2025}}. This framing connects localized patterns of functional centrality to broader questions of urban sustainability without presuming that structural polycentricity has already emerged.

\section{Literature Review}

Urban form organizes the spatial conditions through which activities, infrastructure, and interactions develop. Cities can be understood as coupled urban socioecological systems whose social, spatial, infrastructural, and environmental components interact across scales \textcolor{blue}{\parencite{Batty2005, Bettencourt2013, McPhearson2022, Cioffi2016, Frank2017}}. Within these systems, spatial structure conditions opportunities for interaction, while mobility and everyday activity contribute to differentiated functional configurations. Understanding urban sustainability therefore requires examining not only where urban elements are located, but how they interact.

Research on polycentricity has traditionally identified centers through morphological attributes, including employment, population, built-up density, and economic activity \textcolor{blue}{\parencite{Chen2021, Finka2015}}. These measures characterize the location and hierarchy of structural concentrations, but do not establish whether those concentrations organize meaningful interactions. Functional approaches instead examine flows and relationships: commuting-based and network analyses connect urban organization to travel times and emissions \textcolor{blue}{\parencite{Veneri2010}}, while distinctions between static and dynamic spatial structures emphasize mobility as evidence of how centers operate \textcolor{blue}{\parencite{Wu2021, Thomas2022}}. Morphological and functional centrality are thus related but non-equivalent dimensions of urban organization.

This distinction matters when new infrastructure is introduced into heterogeneous urban contexts. Facilities embedded in established centers may reinforce existing functions, whereas facilities outside those centers may support new functional concentrations when connected to surrounding populations and activities \textcolor{blue}{\parencite{Pacione2009, Pumain2020, Tonkiss2014}}. Multiple localized changes may be consistent with functional restructuring even without an immediate transformation of metropolitan morphology \textcolor{blue}{\parencite{Bramley2009, Echenique2012, Rode2017}}. Their sustainability implications, however, cannot be inferred from polycentricity alone: morphological and functional configurations may differ in their environmental and resilience outcomes \textcolor{blue}{\parencite{Yang2025}}. What matters is whether the resulting organization improves effective proximity, access to opportunities, and the distribution of mobility burdens.

Social infrastructure provides a relevant setting for examining these relationships. It encompasses places, facilities, and institutions supporting collective life, social interaction, and shared services \textcolor{blue}{\parencite{Hall2020, Zahnow2024}}. Its function depends on the populations and interactions it supports, not simply on the provision of buildings or their territorial dispersion \textcolor{blue}{\parencite{Högström2022, Kim2024}}. Consequently, distributing social facilities across underserved areas may expand opportunities without necessarily producing new functional centers. Whether such centers develop depends partly on how infrastructure is embedded within existing urban structures and connected to everyday mobility.

Accessibility offers a way to characterize these connections. Research has moved beyond geometric proximity and density-based measures \textcolor{blue}{\parencite{Gastner2006, Um2009}} toward network-based approaches incorporating travel times, transport modes, and spatially distributed opportunities \textcolor{blue}{\parencite{Geurs2004, Manaugh2012, Karner2018, Xu2020}}. As a relationship between land use and transport, accessibility expresses the ease of reaching activities and services \textcolor{blue}{\parencite{Wachs1973, Geurs2004}}. Multimodal, pedestrian, and distributional analyses further show that these opportunities depend on both network connections and the populations they serve \textcolor{blue}{\parencite{Saif2019, Owen2015, Falchetta2020, Weiss2020}}. Pedestrian connections shape local proximity, while public transport extends the potential reach of facilities across the urban system.

Potential accessibility nevertheless differs from realized function. A well-connected facility may receive limited use, while another may attract substantial activity despite comparatively constrained modeled access. Observed visitation and the spatial origins of users therefore provide complementary evidence of interaction, although visitation alone does not establish centrality. The functional significance of a facility must instead be interpreted in relation to its pre-existing structural context, potential accessibility, and realized territorial reach. Their heterogeneous combinations may indicate reinforcement of established centers or conditions compatible with emerging localized centralities, without demonstrating a dynamic process of urban restructuring.

Multilayer networks are particularly suited to representing these interdependencies. They preserve the distinct organization of pedestrian mobility, public transport services, and social infrastructure while modeling the connections between them \textcolor{blue}{\parencite{Kivela2014}}. Such representations are relevant to coupled socioecological systems, in which system-level outcomes depend on relationships among subsystems rather than on isolated components \textcolor{blue}{\parencite{Synes2019, Sun2025}}. They therefore make it possible to distinguish the network conditions enabling potential interaction from the activity actually observed around individual facilities.

Together, these literatures identify an analytical gap: morphological centrality, functional polycentricity, accessibility, social infrastructure, and multilayer mobility are well studied, but are less often examined jointly to assess how spatially distributed public facilities relate to pre-existing urban structure and realized use. The UTOPÍAS program in Iztapalapa, intended to redistribute access to social infrastructure \textcolor{blue}{\parencite{EVALUA2022}}, provides an opportunity to examine this relationship. By distinguishing structural context, potential accessibility, and realized function, the study evaluates whether facilities reinforce existing centralities or exhibit conditions consistent with emerging functional centers, and what their collective organization can---and cannot---indicate about localized polycentricity and urban sustainability.

\section{Conceptual Framework}

\subsection{Urban form, function, and functional centrality}

The city is understood as a complex urban socioecological system in which spatial structure and human interactions are interdependent. Urban form describes the organization of population, economic activity, infrastructure, and established centralities, providing both an outcome of urban development and the structural context for subsequent interventions \textcolor{blue}{\parencite{Batty2005, Bettencourt2013}}. Urban function, by contrast, concerns the activities and interactions supported by that structure. Their relationship, termed \emph{form--function coupling}, is not fixed: similar spatial configurations may sustain different functions, while new activities may develop outside established centers.

This distinction separates structural from functional centrality. Morphological approaches identify concentrations of population, employment, economic activity, and infrastructure \textcolor{blue}{\parencite{Finka2015, Chen2021, Xu2023}}; functional centrality concerns a place's capacity to attract, organize, and sustain interactions \textcolor{blue}{\parencite{Veneri2010, Wu2021}}. Consequently, structural prominence does not guarantee functional significance, nor must a functionally significant location coincide with an established structural center.

Social infrastructure introduces opportunities for collective activity and access to services, particularly in territories where urban resources are unevenly distributed \textcolor{blue}{\parencite{Hall2020, Zahnow2024, Högström2022, Kim2024}}. Its functional role, however, depends on connections to surrounding populations and activities. Pedestrian and public transport networks condition these connections: accessibility describes the \emph{potential for interaction}, whereas \emph{realized function} refers to interactions evidenced by observed use. Neither infrastructure provision nor accessibility alone establishes functional centrality.

The relationship among pre-existing structure, potential interaction, and realized function permits three comparative interpretations. Substantial function in a structurally prominent location is consistent with \emph{reinforcement}; substantial function in a comparatively weak structural context may indicate an \emph{emerging functional centrality}; and limited integration or realized interaction may indicate \emph{functional constraint}. These are relational interpretations of heterogeneous urban configurations, not intrinsic attributes, predetermined classes of facilities, or evidence of temporal transitions.

\subsection{From functional centrality to localized polycentricity}

The significance of individual centralities for urban form depends on their collective territorial organization. \emph{Localized functional polycentricity} refers to multiple functionally significant centers organizing interactions across different parts of the territory, rather than simply to the spatial dispersion of facilities. Functional change may occur without an immediate morphological transformation; therefore, evidence of emerging localized centralities should not be equated with a demonstrated restructuring of the metropolitan urban system \textcolor{blue}{\parencite{Zhou2024, Thomas2022}}.

The coexistence of reinforced and potentially emerging centralities reflects heterogeneous form--function relationships. Their possible contribution to a more polycentric organization depends on how interactions and opportunities are distributed across the territory, not on the number of facilities alone. This system-level interpretation also avoids treating polycentricity as inherently sustainable \textcolor{blue}{\parencite{Yang2025}}.

Its sustainability implications depend on whether functional organization improves effective proximity, redistributes access to opportunities, and reduces dependence on dominant centers \textcolor{blue}{\parencite{Veneri2010, Yang2025}}. These outcomes depend on interactions among infrastructure, mobility, spatial structure, and human activity within the urban socioecological system \textcolor{blue}{\parencite{Frank2017, Cioffi2016}}. Functional intensity and proximity must nevertheless be distinguished: a facility attracting users from distant locations may be important without improving neighborhood-scale access. Sustainability is therefore discussed in terms of the potential implications of territorial function, rather than evaluated as a directly observed outcome or assumed to follow automatically from decentralization.

\subsection{A multilayer representation of form--function coupling}

A multilayer network representation preserves the distinction between pedestrian mobility, public transport services, and social infrastructure while explicitly modeling their connections \textcolor{blue}{\parencite{Kivela2014}}. Walking links facilities to their immediate surroundings and to transit stops; public transport extends potential reach across the urban system. Accessibility is consequently a relational property of this coupled mobility--infrastructure network, rather than an attribute of a facility's location alone.

Travel times, population coverage, multimodal connectivity, and transfer requirements characterize the conditions for potential interaction \textcolor{blue}{\parencite{Gligorić2023}}. Observed visitation provides separate evidence of realized function, while pre-existing structural indicators establish the urban-form context \textcolor{blue}{\parencite{Xie2026}}. Comparing these dimensions allows heterogeneous form--function configurations to be interpreted without collapsing them into a single measure of centrality \textcolor{blue}{\parencite{Güller2026}}.

\begin{figure}[htbp]
    \centering
    \includegraphics[width=\linewidth]{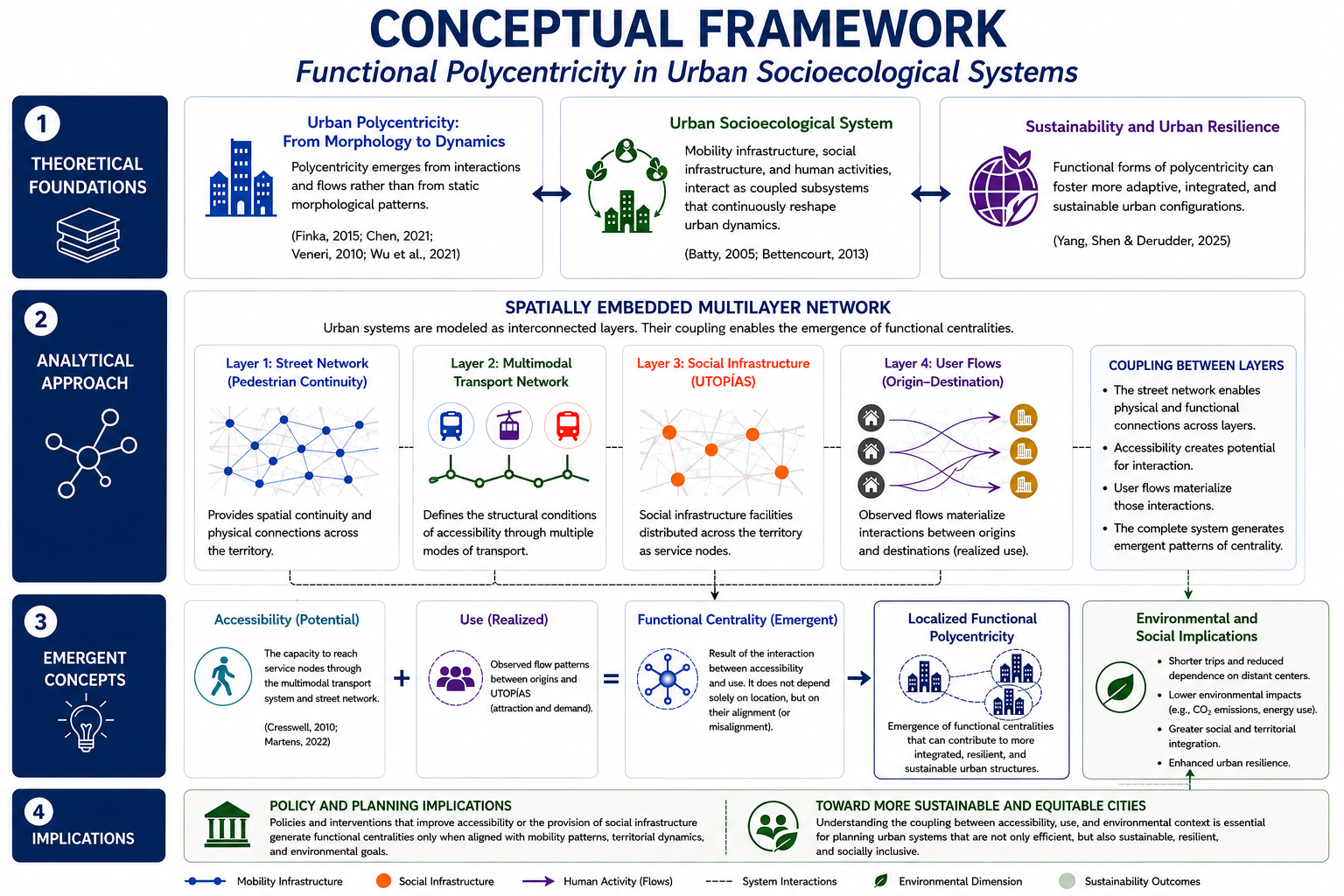}
    \caption{Conceptual framework of urban form--function coupling. Pre-existing urban form provides the structural context in which social infrastructure is embedded. Pedestrian and multimodal mobility systems generate conditions of potential interaction through accessibility, while realized interactions characterize effective urban function. Comparing these dimensions supports the interpretation of configurations consistent with reinforced, potentially emerging, or functionally constrained centralities, without establishing temporal transitions or causal effects. The territorial organization of multiple functionally significant centralities may contribute to localized functional polycentricity, whose sustainability implications depend on proximity, redistribution of opportunities, and social inclusion.}
    \label{fig:conceptual_framework}
\end{figure}

Figure~\ref{fig:conceptual_framework} summarizes these relationships. It links structural context, network-enabled opportunities, and realized interactions to the possible reinforcement or emergence of localized functional centralities. The diagram represents an analytical relationship, not a deterministic sequence or evidence that a system-wide transition has already occurred.

\subsection{Operationalizing the urban form--function relationship}

The framework is operationalized through three distinct analytical dimensions. \emph{Pre-existing urban form} is represented by the Pre-existing Centrality Index (PCI) and its neighborhood-weighted extension (wPCI), independently of subsequent UTOPÍA use. \emph{Potential interaction} is characterized by pedestrian and multimodal travel-time accessibility, population coverage, and transfer conditions. \emph{Realized function} is assessed through observed visitor intensity, home-to-site distances, and the territorial distribution of user origins. Visitors' socioeconomic characteristics add a distributional dimension but are not treated as a centrality score.

The comparison focuses on convergence and divergence among these dimensions. For example, high modeled accessibility with limited observed use indicates unrealized potential, while substantial use despite weaker modeled accessibility suggests functions not fully captured by formal transport connectivity. Substantial use in an already strong structural context is interpreted differently from comparable use outside established centralities. Such contrasts inform interpretations of reinforcement, possible emergence, and functional constraint without establishing causal effects of infrastructure provision.

Table~\ref{tab:conceptual_operationalization} records the conceptual and operational roles of these measures. They are neither interchangeable nor combined automatically into a single index; their relationships provide the basis for assessing localized functional configurations and, collectively, possible functional polycentricity.

\begin{table*}[htbp]
\centering
\caption{Conceptual and operational dimensions used to evaluate functional centrality and localized polycentricity.}
\label{tab:conceptual_operationalization}
\small
\setlength{\tabcolsep}{5pt}
\renewcommand{\arraystretch}{1.15}

\begin{tabularx}{\textwidth}{
>{\raggedright\arraybackslash}p{3.0cm}
>{\raggedright\arraybackslash}p{4.2cm}
>{\raggedright\arraybackslash}X
}
\toprule
\textbf{Conceptual Dimension} & \textbf{Operational Indicator} & \textbf{Interpretation} \\
\midrule

Pre-existing Urban Form & PCI and wPCI & Structural centrality and surrounding spatial context independently from the functional role subsequently developed by UTOPÍAs. \\

Potential Interaction & Population coverage and multimodal travel-time accessibility & Opportunities for interaction and proximity generated through the integration of each UTOPÍA with the pedestrian and public transport system. \\

Potential Interaction & Multimodal connectivity and transfer conditions & Degree of integration of each facility within the coupled mobility system and the travel frictions associated with reaching it. \\

Realized Function & Visitor intensity & Magnitude of observed activity associated with each UTOPÍA. \\

Realized Function & Spatial distribution and distance of user origins & Territorial reach, effective catchment, and spatial scale of realized interaction. \\

Distributional Sustainability & Socioeconomic characteristics of users & Social distribution of realized access, interpreted as an equity dimension rather than as centrality. \\

Form--Function Coupling & Relationship among pre-existing centrality, potential interaction, and realized function & Evidence of convergence or divergence between structural position, accessibility conditions, and effective urban function. \\

Functional Centrality & Reinforcement, emergence, or functional constraint & Functional role of each UTOPÍA relative to the pre-existing urban structure. \\

Localized Functional Polycentricity & Territorial organization of multiple functional centralities & Evidence of multiple functionally significant centers and their organization within the study area. \\

\bottomrule
\end{tabularx}
\end{table*}

This operationalization connects the conceptual framework to the empirical analysis. The structural indicators define the baseline, the multilayer model characterizes potential interactions, and independent visitation observations describe realized function. Their heterogeneous relationships can identify conditions compatible with emerging centralities, while claims about structural polycentricity or sustainability outcomes require evidence beyond the presence and accessibility of facilities.

\clearpage

\section{Methods}

\subsection{Study area and data sources}\label{sec:methods-data}

The study examines 15 UTOPÍAs in Iztapalapa, Mexico City: Olini, Meyehualco, Barco Utopía, Libertad, Quetzalcóatl, Tezontli, Papalotl, Atzintli, Teotongo, Tecoloxtitlan, La Cascada, Cuauhtlicalli, Ixtapalcalli, Estrella Huizachtépetl (Cahui), and Cihuacóatl. Iztapalapa defines the study's territorial and policy focus, but public transport services are not truncated at its administrative boundary. The facilities are analyzed as social infrastructure destinations, not presumed urban centers.

\begin{figure}
    \centering
    \includegraphics[width=0.85\linewidth]{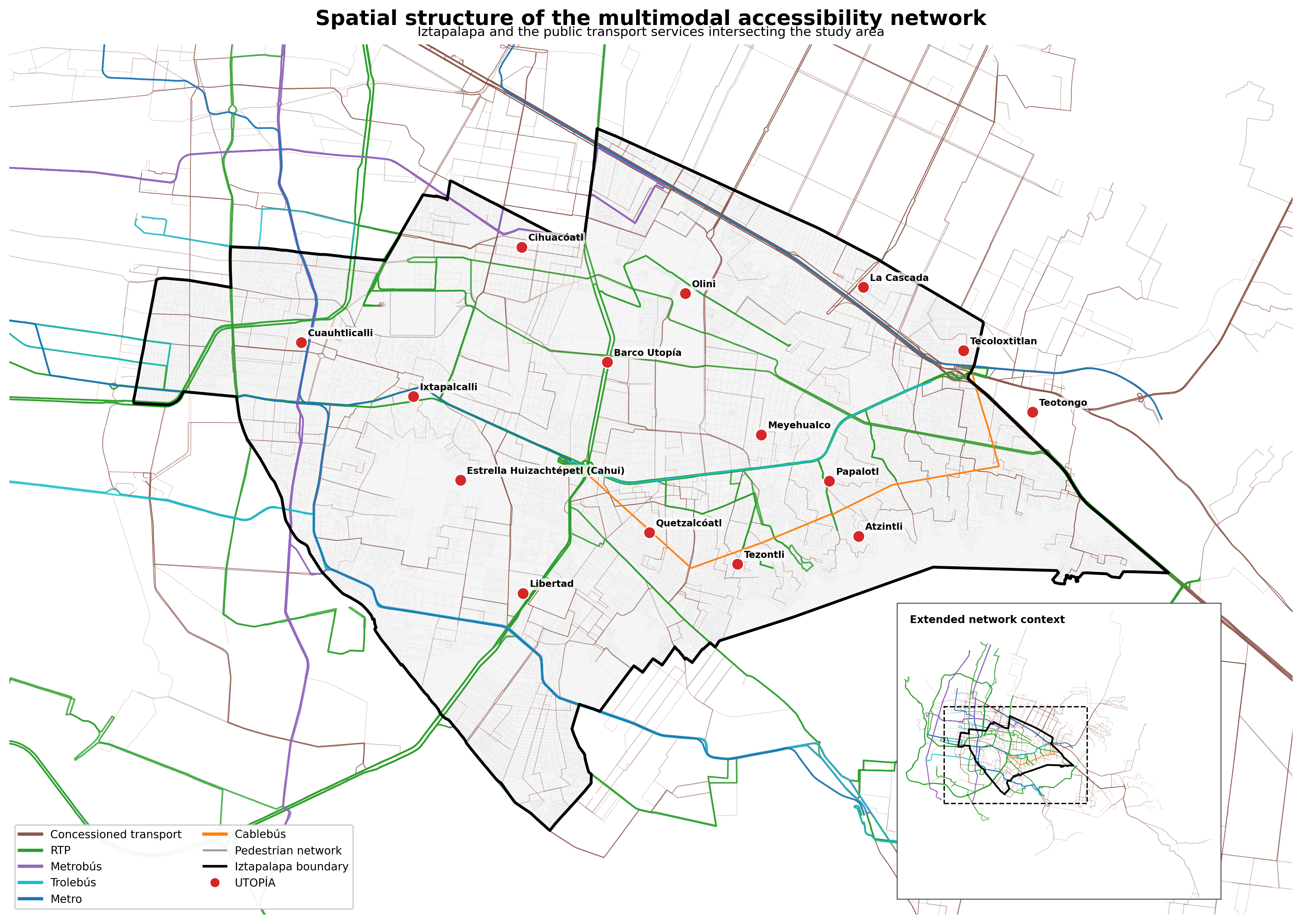}
    \caption{Spatial configuration of the multimodal accessibility network in Iztapalapa, Mexico City. The map illustrates the spatial distribution of the 15 UTOPÍAS and the public transport services from a Mexico City GTFS file \textcolor{blue}{\parencite{WIMT2024}} intersecting the study area, including Metro, Metrobús, Trolebús, Cablebús, RTP, and concessioned transport. The pedestrian street network provides the spatial basis for integrating the different transport modes and connecting them to the UTOPÍAs. The black boundary delineates Iztapalapa, while the inset shows the broader spatial extent of the transport network considered in the analysis.}
    \label{fig:redgthfs}
\end{figure}

The analysis combines georeferenced UTOPÍA locations; economic establishment abundance and diversity; population and physical public transport stops aggregated to H3 hexagons; a topologically corrected pedestrian street network; and Mexico City General Transit Feed Specification (GTFS) data, including concessioned transport \parencite{WIMT2024}. Independent location-based visitation data provide evidence of realized function \parencite{Langle2026}. The analytical workflow comprises five stages: data assembly; spatial and GTFS preprocessing; multilayer network construction; accessibility computation; and scenario-based comparisons integrating population, structural indicators, and observed use. The corresponding outputs are reported in the Results section. Figure~\ref{fig:redgthfs} shows the geographical scope of the selected mobility services.

\subsection{Pre-existing urban structure}\label{sec:methods-structure}

The characterization of pre-existing urban structure draws on multidimensional approaches to urban centrality that consider the spatial concentration and diversity of economic activities, population distribution, and transport infrastructure \textcolor{blue}{\parencite{Wen2024, Knaap2024, Güller2026}}. Building on this literature, structural centrality was characterized on H3 resolution-8 (R8) hexagons using four variables: economic establishment abundance, economic diversity, resident population, and the number of distinct physical GTFS stops. Counting physical stops, rather than service-pattern nodes, avoids multiplying infrastructure at locations served by several routes. Each variable was min--max normalized, and the four normalized values were averaged with equal weights to obtain the Pre-existing Centrality Index (PCI). Thus, PCI provides a composite characterization of local structural conditions rather than directly identifying urban subcenters or measuring modeled accessibility and observed UTOPÍA use.

Because local centrality is embedded in a broader spatial context, the structural characterization also considers the surrounding urban environment, consistent with approaches that emphasize the relative importance of centers and their spatial relationships \textcolor{blue}{\parencite{Jacobs2023, Knaap2024}}. Each UTOPÍA was assigned the PCI of its containing hexagon. Its surrounding context was measured using neighborhood-weighted PCI (wPCI), calculated from the 12 nearest hexagons excluding the focal cell, with inverse-distance weights (\(k=12\), distance-decay exponent \(p=1\)). The alternative neighborhood specifications and mathematical definitions are documented in Appendix~B. PCI and wPCI are therefore interpreted as complementary measures of local and neighboring structural conditions, whose correspondence with multimodal accessibility and independently observed visitation is examined empirically rather than assumed.

The final structural analysis encompasses all 15 UTOPÍAs, including Teotongo following correction of its geographic assignment. The H3 R8 cells used for population-weighted origins (described below) should not be interpreted as necessarily identical to the complete spatial support of the structural indicator calculations.

\subsection{GTFS-based multimodal network and accessibility}\label{sec:methods-accessibility}

\begin{figure}
    \centering
    \includegraphics[width=1\linewidth]{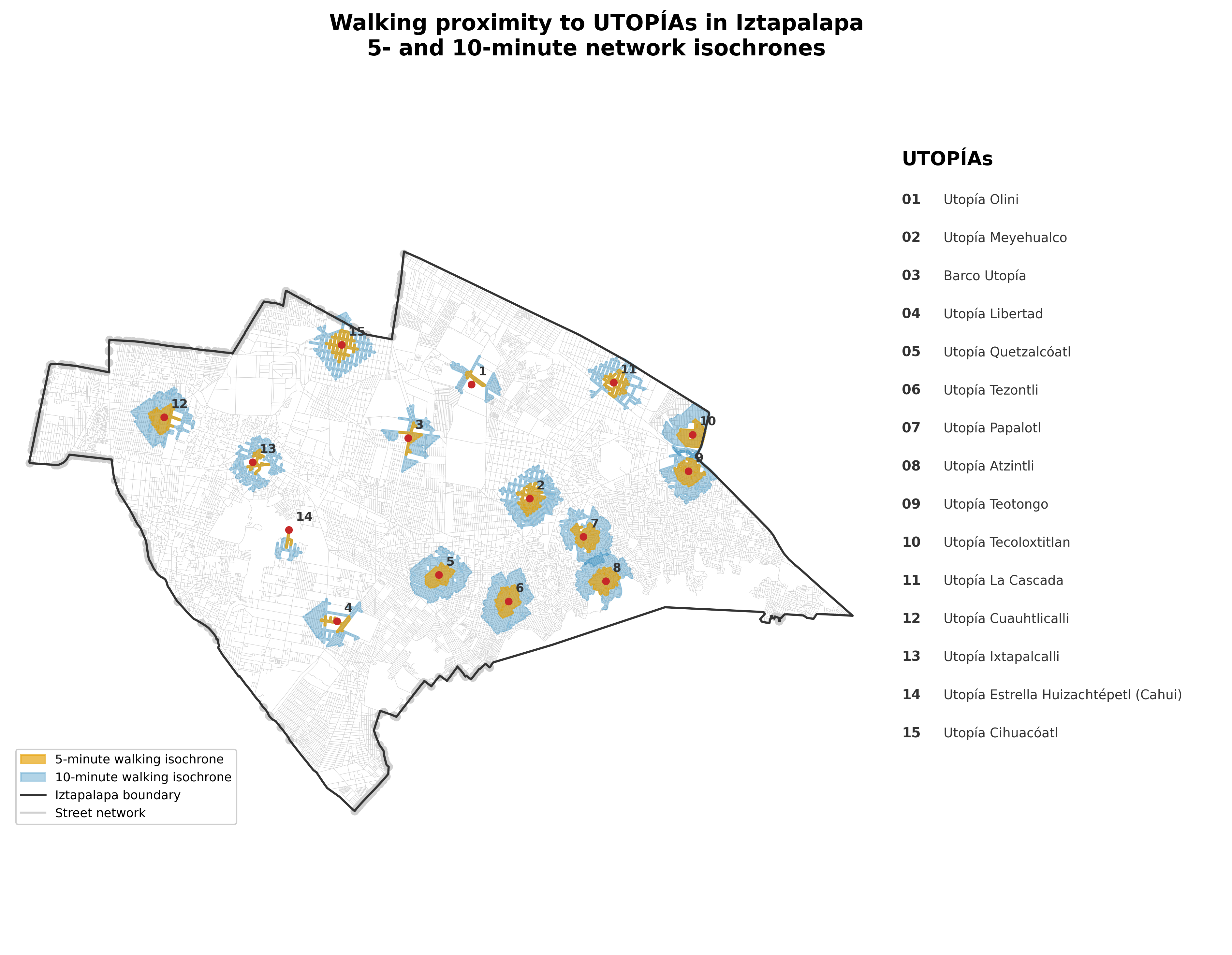}
    \caption{Walking accessibility to the 15 UTOPÍAs in Iztapalapa, Mexico City. Spatial distribution of 5- and 10-minute walking isochrones derived from the pedestrian street network. Orange and blue areas represent locations within the corresponding walking-time thresholds of each UTOPÍA, while the gray street network provides spatial context. The black outline delineates the administrative boundary of Iztapalapa, and numbered red markers identify the 15 UTOPÍAs listed in the accompanying legend. The figure illustrates the spatial extent and distribution of local pedestrian accessibility to social infrastructure, providing a proximity-based reference for comparison with the multimodal accessibility results.}
    \label{fig:isocronas}
\end{figure}

The mobility model couples a pedestrian street graph, GTFS-based public transport services, and UTOPÍA destination nodes \parencite{Kivela2014}. The corrected pedestrian graph comprises 24,108 intersection or segment-endpoint nodes and 67,498 directed walking edges weighted by travel time. It serves both as the walk-only network and as the spatial interface between origins, public transport stops, and facilities. Five- and ten-minute walking isochrones provide local proximity measures (Figure~\ref{fig:isocronas}).

\begin{figure}
    \centering
    \includegraphics[width=1\linewidth]{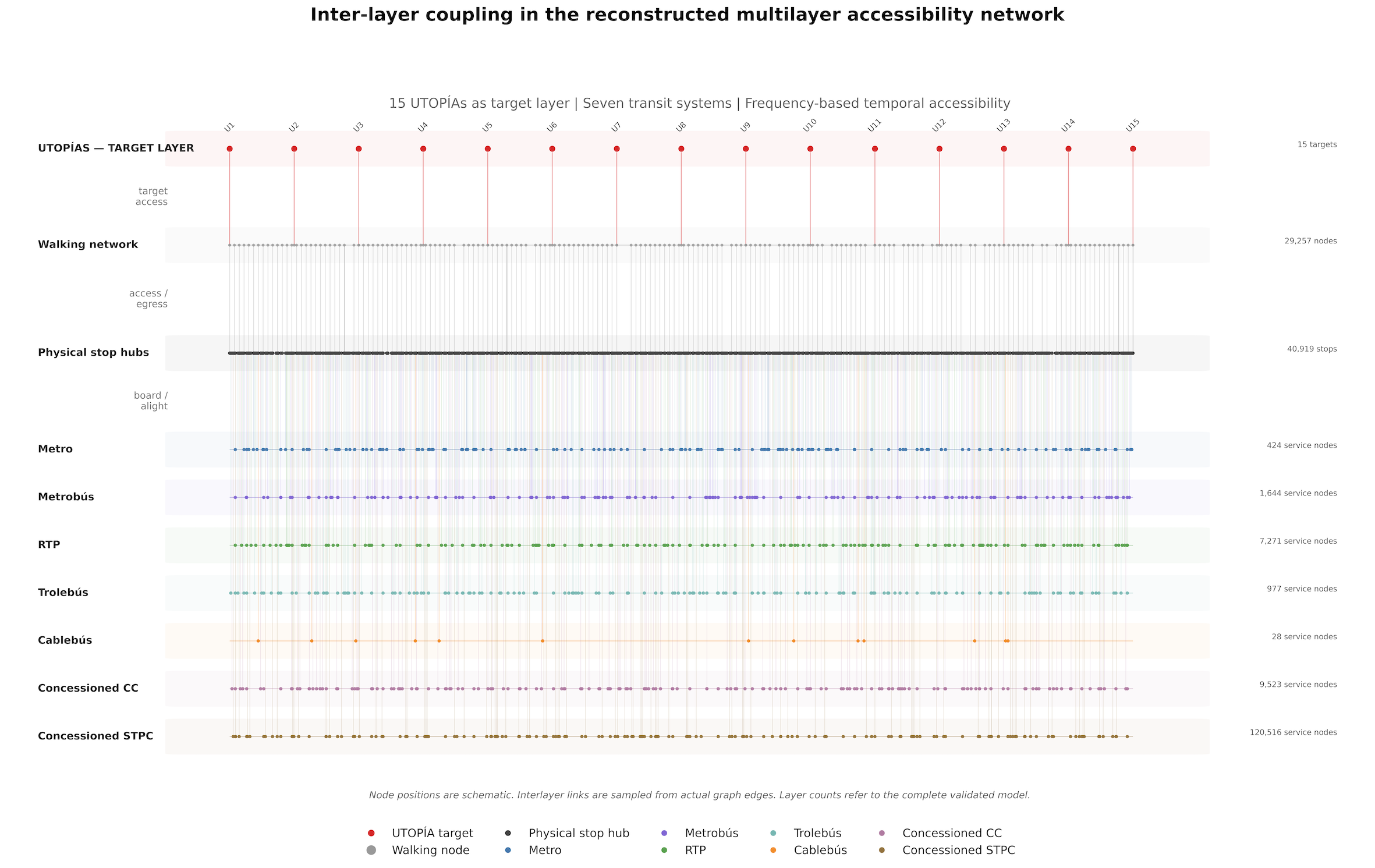}
    \caption{Inter-layer coupling in the reconstructed GTFS-based multilayer accessibility network. The diagram illustrates the integration of the 15 UTOPÍAs, the pedestrian street network, physical-stop hubs, and public transport service layers, including Metro, Metrobús, RTP, Trolebús, Cablebús, and concessioned transport. Physical-stop hubs provide the interface between pedestrian access and route-specific service nodes, enabling boarding, alighting, and transfers between transport services. Walking connections link the different mobility components to the UTOPÍAs, which are represented as destination nodes rather than intermediate transfer points. This architecture preserves the operational distinction between transport services while enabling the calculation of multimodal travel-time accessibility.}
    \label{fig:ML_1}
\end{figure}

GTFS routes were selected if at least one stop served Iztapalapa, while the full stops and trajectories of those services were retained beyond the borough. The selected feed comprises 386 routes, 1,663 trips, and 10,568 physical stops (3,865 inside Iztapalapa and 6,703 outside). It covers Metro, Metrobús, RTP, trolleybus, Cablebús, light rail, and concessioned transport. In the GTFS-HUB representation, geographic physical-stop hubs connect walking links to route-specific service-pattern nodes. Boarding adds expected waiting time, service edges encode in-vehicle time, and alighting returns passengers to a physical hub; transfers occur through hubs or walking links between stops. The primary temporal scenario is Wednesday at 12:00, with services selected using GTFS calendar and frequency information. Of the selected patterns, 813 were active on the representative weekday and 776 frequency patterns at noon. For frequency-based services, expected wait equals half the headway (median headway 10 minutes; median expected wait 5 minutes). Each UTOPÍA is connected to the pedestrian graph by a final walking link and is treated exclusively as a target: no UTOPÍA-to-UTOPÍA or intermediate-transfer links are permitted (Figure~\ref{fig:ML_1}).

Travel times were estimated by shortest paths on the time-weighted graph using Dijkstra's algorithm. Depending on the route, total time comprises origin access, walking, expected waiting, in-vehicle travel, transfer, egress, and final destination access:
\[
T_{od}=T_{\mathrm{origin}}+T_{\mathrm{walk}}+T_{\mathrm{wait}}+T_{\mathrm{transit}}+T_{\mathrm{transfer}}+T_{\mathrm{egress}}+T_{\mathrm{target}}.
\]
The decomposition permits reporting total travel time alongside its component costs and the number of boardings or transfers; walk-only paths have no public transport components.

\begin{figure}
    \centering
    \includegraphics[width=.9\linewidth]{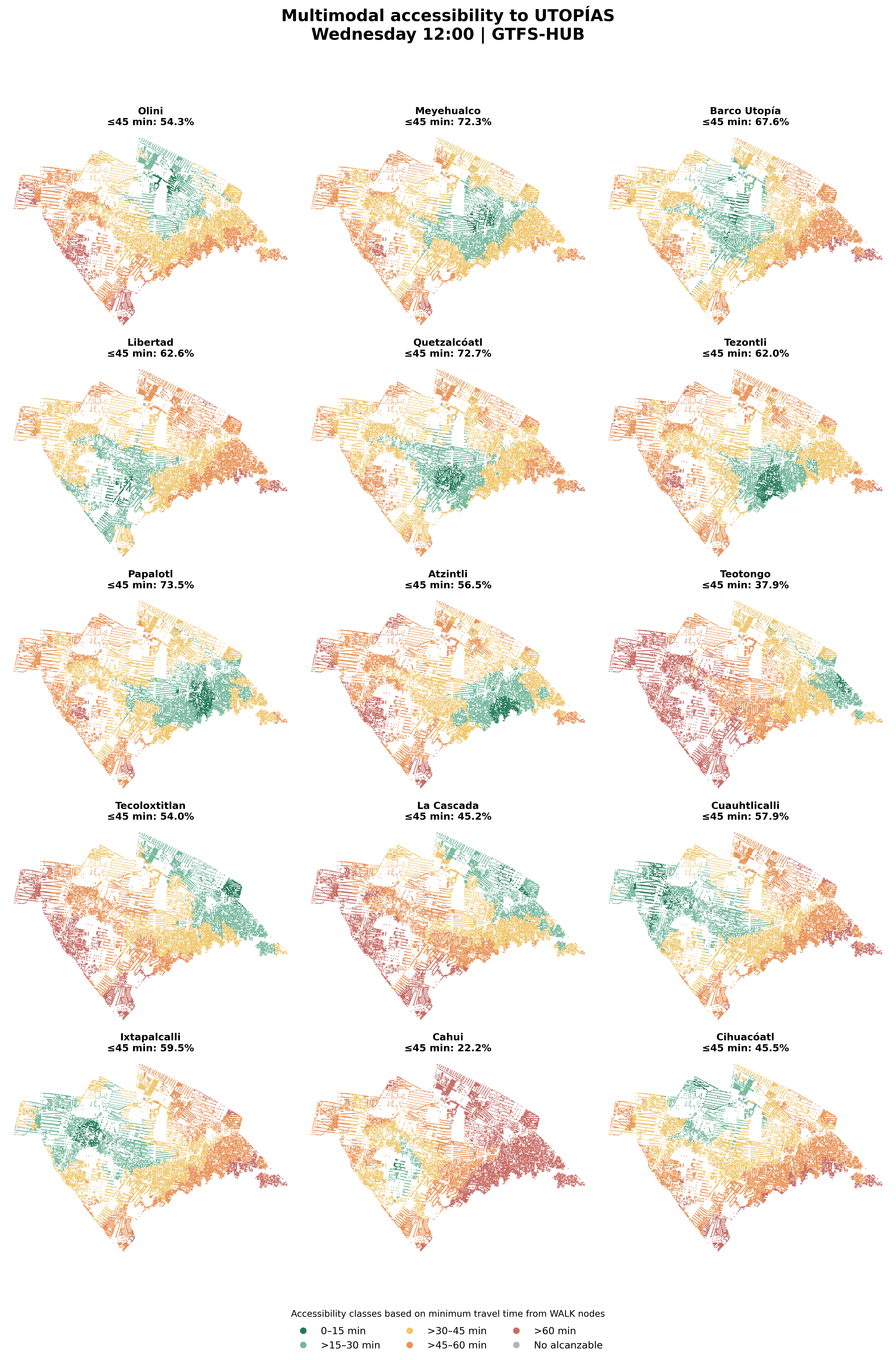}
    \caption{Multimodal accessibility to the 15 UTOPÍAs in Iztapalapa. Spatial distribution of minimum inbound travel times from pedestrian network nodes to each UTOPÍA, estimated using the frequency-based GTFS-HUB multimodal network for Wednesday at 12:00. Each panel represents an individual UTOPÍA, with travel times classified into five intervals (0–15, >15–30, >30–45, >45–60, and >60 minutes), alongside unreachable nodes. Percentages indicate the proportion of evaluated pedestrian nodes able to reach the corresponding UTOPÍA within 45 minutes. The maps illustrate spatial differences in potential multimodal accessibility across the 15 facilities; they do not represent observed visitation or population coverage.}
    \label{fig:MMAU}
\end{figure}

Two complementary spatial representations were used for accessibility. First, pedestrian intersection nodes were evaluated as network origins or destinations to map minimum inbound times to each UTOPÍA, temporal reachability profiles, and spatially differentiated effects of transport services (Figure~\ref{fig:MMAU}). H3 R9 hexagons were used as a higher-resolution spatial representation of accessibility outputs; they are distinct from pedestrian routing nodes and the population-weighted H3 R8 origins. The spatial assignment of node-based travel times to R9 cells is a representation step, not a separate routing model. Second, population-weighted accessibility was calculated from 162 H3 R8 origin cells, representing approximately 1.83 million residents, to the 15 UTOPÍAs. Each origin was connected to the walking network before multimodal routing. One origin (11,786 residents) was disconnected from every facility and was retained in coverage denominators. The resulting origin--destination matrix contains 2,430 possible pairs, of which 2,415 are reachable.

Accessibility indicators include population-weighted median and 90th-percentile travel times, walk-only trip share, journey components, and the proportion of residents reaching each UTOPÍA within 45 minutes:
\[
C_{45,j}=\frac{\sum_i P_i\,\mathbb{I}(T_{ij}\leq45)}{\sum_i P_i},
\]
where \(P_i\) is origin population and \(T_{ij}\) is the minimum modeled travel time to facility \(j\). The denominator includes all 162 origin cells. Four complementary analytical scenarios organize the comparisons. In the baseline multimodal scenario, the full walking--GTFS-HUB network was used to compute minimum travel times to each of the 15 UTOPÍAs, reachability of pedestrian nodes at 10-, 20-, 30-, and 45-minute thresholds, mapped accessibility patterns, and population-weighted coverage from H3 R8 origins. The five- and ten-minute pedestrian isochrones were retained as a separate walk-only proximity reference. In the mode-exclusion scenario, each public transport system was removed in turn while the remaining network was retained; changes in minimum travel time and spatial reachability relative to the baseline describe each system's modeled marginal contribution. In the directional-accessibility scenario, inbound journeys to UTOPÍAs were compared with outbound journeys from them using the directed, time-weighted network, permitting assessment of travel-time and reachability asymmetry. Finally, the population-and-functional-context scenario related baseline accessibility and coverage indicators to PCI/wPCI and independently observed visitation measures. These comparisons characterize potential accessibility, system dependence, directional differences, and form--function relationships; they do not establish causal effects of infrastructure provision or observed transport ridership.

\subsection{Realized function and form--function analysis}\label{sec:methods-function}

Realized function was characterized using independent location-based mobility observations \parencite{Langle2026}, not official facility attendance records. Variables comprise observed visitor counts (log-transformed for intensity comparisons), median home-to-site distance, the percentage of observed visitors residing in Iztapalapa, and socioeconomic attributes of visitors. The latter inform distributional interpretation rather than a centrality score. Visitation data were available for 13 UTOPÍAs; Estrella Huizachtépetl and Cihuacóatl lack these observations. The final structural dataset includes Teotongo. Accordingly, sample sizes for associations combining structural indicators and visitation must be established from the complete-case observations used in each test, rather than inferred from the earlier geographic exclusion of Teotongo. The GTFS-based accessibility model and the independent visitation dataset represent different observation processes and are not assumed to describe temporally equivalent conditions. Consequently, their associations are interpreted as exploratory relationships between modeled accessibility and observed use, rather than as direct evidence of behavioral responses to the modeled transport conditions.

Six pairwise associations were specified: PCI versus the share of visitors from Iztapalapa; wPCI versus median home-to-site distance; 45-minute population coverage versus log visitor counts; population-weighted median travel time versus log visitor counts; walk-only accessibility share versus log visitor counts; and wPCI versus the population-weighted 90th-percentile travel time. Pearson's \(r\) and Spearman's \(\rho\) were reported, with 10,000-permutation tests, 5,000-bootstrap confidence intervals, and leave-one-out sensitivity analyses to assess the stability of associations in these small samples. The number of valid observations is reported separately for each association, reflecting the available variables.

Rather than combining structural, accessibility, and visitation measures into a single centrality score, the analysis interprets their convergence and divergence. The comparison distinguishes structural concentration, modeled opportunities for access, and observed territorial interactions without assuming that these dimensions coincide. Any interpretation of reinforced, emerging, or constrained functional centrality is therefore conditional on the empirical evidence, not a mechanically assigned or causally identified classification. Localized functional polycentricity is assessed as a possible pattern of distributed functional relationships rather than a demonstrated restructuring of the urban system.

\clearpage

\section{Results}

Our integrated analysis reveals that the relationship between urban form and realized function cannot be reduced to either pre-existing centrality or multimodal accessibility alone. Instead, the UTOPÍAs exhibit heterogeneous configurations in which structural context, accessibility conditions, and realized spatial interactions combine differently across facilities.

\begin{table}[htbp]
\centering
\caption{Average contribution of the main components to multimodal travel time. The general pattern shows that accessibility to UTOPÍAs is clearly multimodal and dependent on public transport. On average, time spent in transit accounts for 44.3\% of the journey and is the dominant component at 14 of the 15 facilities. However, walking remains highly significant (30.8\%), followed by waiting (17.2\%). Together, walking and waiting make up nearly 48\% of the average time, indicating that accessibility depends not only on the speed of the transport network but also on its pedestrian connectivity and service frequency. This aligns well with our interpretation of mobility as a mechanism mediating the relationship between social infrastructure and urban function.}
\label{tab:travel_time_components}
\begin{tabular}{lcc}
\hline
\textbf{Travel-time component} & \textbf{Mean share (\%)} & \textbf{Dominant in UTOPÍAs} \\
\hline
In-vehicle public transport & 44.3 & 14 of 15 \\
Walking                     & 30.8 & 1 of 15  \\
Waiting                     & 17.2 & 0 of 15  \\
\hline
\textbf{Combined share}      & \textbf{92.3} & -- \\
\hline
\end{tabular}
\begin{flushleft}
\footnotesize
\textit{Note:} Shares correspond to the mean across the 15 UTOPÍAs of population-weighted travel-time components. The remaining 7.7\% corresponds to origin access, stop access, egress, and final pedestrian connection to the UTOPÍA.
\end{flushleft}
\end{table}

\subsection{Accessibility structure of the transport network}
The analysis characterizes the structural and temporal conditions of accessibility generated by the integrated public transport network in Iztapalapa. The multimodal model incorporates Metro, Metrobús, Cablebús, Trolebús, RTP, and concessioned transport services represented in the General Transit Feed Specification (GTFS) datasets, together with their connections to the pedestrian street network. By integrating these systems within a unified multilayer framework, the analysis captures spatial variations in network connectivity, service availability, and potential travel times across the territory (Table~\ref{tab:travel_time_components}) and (Table~\ref{tab:utopias_stops_isochrones}). These conditions provide the basis for examining how the transport network shapes access to UTOPÍAs and the spatial organization of potential functional centralities (Figure \ref{fig:redgthfs}).

Rather than relying on a binary distinction between 'formal' and 'informal' public transport, the model adopts an operational criterion based on the availability of spatially and temporally structured transport data. The integrated GTFS-based representation incorporates routes, stops, service frequencies, and transfer connections, enabling the comparison of different transport systems within a consistent analytical framework. However, the resulting accessibility estimates reflect the operational assumptions encoded in the available datasets and do not fully capture day-to-day service variability, irregular operations, overcrowding, perceived safety, or other experiential conditions affecting actual mobility. Accordingly, inclusion in the model should not be interpreted as evidence of equivalent service reliability, institutional coordination, or operational performance across transport systems.

From a sustainable mobility perspective, this integrated approach enables the analysis of how different public transport systems collectively and differentially contribute to accessibility to social infrastructure. Incorporating both institutionally operated and concessioned services provides a more comprehensive representation of potential mobility opportunities and allows the effects of individual transport systems on modeled travel times to be examined (Table~\ref{tab:utopias_transport_access}). Nevertheless, greater modeled connectivity or shorter travel times do not necessarily translate into more equitable accessibility, reliable service provision, or sustainable mobility outcomes. These outcomes also depend on socioeconomic, institutional, and operational conditions beyond the scope of the network model. In this context, fragmentation and limited coordination among mobility services may introduce spatial and temporal uncertainties that constrain the development of more sustainable and functionally integrated urban mobility systems \textcolor{blue}{\parencite{Banister2005}}.

\begin{table}[p]
\centering
\caption[Pedestrian access to public transport at UTOPÍAs]{Nearest public transport stops and pedestrian access times by transport system category across the 15 UTOPÍAs in Iztapalapa.}
\label{tab:utopias_transport_access}
\begingroup
\fontsize{8}{9.5}\selectfont
\setlength{\tabcolsep}{2.5pt}
\renewcommand{\arraystretch}{1.14}
\resizebox{\textwidth}{!}{%
\begin{tabular}{@{}>{\RaggedRight\arraybackslash}p{2.8cm} >{\RaggedRight\arraybackslash}p{2.6cm} >{\centering\arraybackslash}p{1.15cm} >{\RaggedRight\arraybackslash}p{5.0cm} >{\centering\arraybackslash}p{1.15cm} >{\RaggedRight\arraybackslash}p{5.0cm} >{\centering\arraybackslash}p{1.15cm}@{}}
\toprule
\textbf{UTOPÍA} & \multicolumn{2}{c}{\textbf{Metro / Metrobús}} & \multicolumn{2}{c}{\textbf{RTP}} & \multicolumn{2}{c}{\textbf{Concessioned transport}} \\
\cmidrule(lr){2-3}\cmidrule(lr){4-5}\cmidrule(l){6-7}
& \textbf{Nearest station} & \textbf{Walk (min)} & \textbf{Nearest stop} & \textbf{Walk (min)} & \textbf{Nearest stop} & \textbf{Walk (min)} \\
\midrule
Olini & Guelatao & 26.89 & Prolongación Plutarco Elías Calles -- C.-S/C C. Perdida & 2.82 & Plutarco Elías Calles Av. Guelatao & 6.12 \\
Meyehualco & Peñón Viejo & 42.98 & Calzada Ermita Iztapalapa -- Reforma Urbana & 9.43 & Av. Circunvalación & 1.30 \\
Barco Utopía & Constitución de 1917 & 28.36 & Anillo Periférico Canal De Garay -- Eje 6 Sur Luis Méndez & 2.53 & Calle 4 Periférico Canal de San Juan & 1.79 \\
Libertad & Periférico Oriente & 13.16 & Anillo Periférico Canal De Garay -- Reclusorio Ote. & 3.27 & Puente Peatonal Prol. Anillo Periférico & 2.94 \\
Quetzalcóatl & Constitución de 1917 & 30.28 & Avenida Insurgentes -- Altamirano & 9.80 & Tienda de Abarrotes Calle Ignacio Manuel Altamirano & 4.65 \\
Tezontli & Olivos & 50.48 & Camino Real -- Pozos & 4.16 & Trigal & 1.06 \\
Papalotl & Acatitla & 44.55 & Camino A Las Minas -- Tejocote & 1.76 & C, Tejocote & 1.59 \\
Atzintli & Acatitla & 54.97 & Camino A Las Minas -- Colonia Palmas Calle Las Margaritas & 6.21 & 2da. Cda. Uvas & 2.13 \\
Teotongo & Santa Marta & 14.16 & Calzada Ermita Iztapalapa -- Luis Cervantes & 11.41 & Calle Tetepetongo & 1.43 \\
Tecoloxtitlan & Santa Marta & 4.69 & Terminal Cetram Metro Santa Martha & 4.77 & Tienda León & 1.93 \\
La Cascada & Peñón Viejo & 14.06 & Ejército De Oriente & 24.82 & Eugenio Aguirre & 3.51 \\
Cuauhtlicalli & Escuadrón 201 & 3.91 & Circuito Interior -- Eje 3 Oriente Francisco Del Paso y Troncoso & 8.30 & Calle Teniente Javier Martínez & 2.60 \\
Ixtapalcalli & Iztapalapa & 8.48 & Calzada Ermita Iztapalapa -- Estrella & 2.19 & Estrella Calz. Ermita Iztapalapa & 2.19 \\
Estrella Huizachtépetl (Cahui) & Calle 11 & 41.76 & Anillo Periférico Canal De Garay -- San Nicolas & 34.93 & Calle Cerezo & 7.80 \\
Cihuacóatl & Canal de San Juan & 37.70 & Eje 5 Sur Av. Leyes De Reforma -- Rojo Gómez & 11.42 & Calle 11 de Agosto de 1859 \& Calle Batalla de Casa Blanca & 0.76 \\
\bottomrule
\end{tabular}%
}
\par\vspace{3pt}
\begin{minipage}{\textwidth}
\footnotesize\textit{Note:} Walking times represent the shortest pedestrian travel time from each UTOPÍA to the nearest accessible stop within each transport category, calculated using the integrated GTFS-HUB multimodal network. Metro and Metrobús are grouped as high-capacity transit systems; RTP is reported separately; and concessioned transport includes Concessioned Transport and Concessioned Corridors services. Walking times exclude waiting and in-vehicle travel times.
\end{minipage}
\endgroup
\end{table}

\begin{table}[htbp]
\centering
\caption{Public transport stops within 5- and 10-minute pedestrian isochrones around the 15 UTOP\'{I}As in Iztapalapa.}
\label{tab:utopias_stops_isochrones}
\begin{minipage}{\textwidth}
\centering
\fontsize{7.5}{9}\selectfont
\setlength{\tabcolsep}{2pt}
\renewcommand{\arraystretch}{1.20}
\resizebox{\textwidth}{!}{%
\begin{tabular}{@{}l*{10}{c}@{}}
\toprule
\multirow{2}{*}{\textbf{UTOP\'{I}A}} & \multicolumn{2}{c}{\textbf{Metro / Metrob\'{u}s}} & \multicolumn{2}{c}{\textbf{RTP}} & \multicolumn{2}{c}{\textbf{Concessioned}} & \multicolumn{2}{c}{\textbf{Cableb\'{u}s}} & \multicolumn{2}{c}{\textbf{Trolleybus}} \\
\cmidrule(lr){2-3}\cmidrule(lr){4-5}\cmidrule(lr){6-7}\cmidrule(lr){8-9}\cmidrule(l){10-11}
& \textbf{5 min} & \textbf{10 min} & \textbf{5 min} & \textbf{10 min} & \textbf{5 min} & \textbf{10 min} & \textbf{5 min} & \textbf{10 min} & \textbf{5 min} & \textbf{10 min} \\
\midrule
Olini & 0 & 0 & 3 & 5 & 0 & 21 & 0 & 0 & 0 & 0 \\
Meyehualco & 0 & 0 & 0 & 2 & 3 & 29 & 0 & 0 & 0 & 1 \\
Barco Utop\'{i}a & 0 & 0 & 3 & 7 & 7 & 28 & 0 & 0 & 0 & 0 \\
Libertad & 0 & 0 & 3 & 5 & 7 & 43 & 0 & 0 & 0 & 0 \\
Quetzalc\'{o}atl & 0 & 0 & 0 & 2 & 4 & 37 & 0 & 0 & 0 & 0 \\
Tezontli & 0 & 0 & 2 & 10 & 15 & 46 & 0 & 0 & 0 & 0 \\
Papalotl & 0 & 0 & 4 & 12 & 11 & 44 & 0 & 0 & 0 & 1 \\
Atzintli & 0 & 0 & 0 & 8 & 13 & 45 & 0 & 0 & 0 & 0 \\
Teotongo & 0 & 0 & 0 & 0 & 18 & 53 & 0 & 0 & 0 & 0 \\
Tecoloxtitlan & 1 & 1 & 0 & 0 & 8 & 41 & 0 & 1 & 0 & 0 \\
La Cascada & 0 & 0 & 0 & 0 & 4 & 31 & 0 & 0 & 0 & 0 \\
Cuauhtlicalli & 3 & 5 & 0 & 2 & 15 & 37 & 0 & 0 & 0 & 0 \\
Ixtapalcalli & 0 & 4 & 3 & 11 & 15 & 69 & 0 & 0 & 0 & 0 \\
Estrella Huizacht\'{e}petl (Cahui) & 0 & 0 & 0 & 0 & 0 & 2 & 0 & 0 & 0 & 0 \\
Cihuac\'{o}atl & 1 & 2 & 0 & 0 & 5 & 52 & 0 & 0 & 0 & 0 \\
\bottomrule
\end{tabular}
}
\par\vspace{0.25cm}
{\footnotesize\raggedright\textit{Note:} Values are counts of distinct physical GTFS-HUB stops intersecting the buffered pedestrian isochrone polygons (5 and 10 minutes; 25-m buffer). Stops serving multiple routes are counted once per transport category. Concessioned combines Concessioned Transport and Concessioned Corridors. The polygons were generated independently and the 5-minute polygons are not fully contained in the 10-minute polygons; consequently, strict geometric nesting has not been established.\par}
\end{minipage}
\end{table}

The spatial distribution of multimodal accessibility reveals an uneven territorial pattern, in which major public transport corridors and intermodal connections are associated with greater potential accessibility to social infrastructure (Figure \ref{fig:scatter_p}). This spatial configuration highlights how the integration of transport systems can extend accessibility beyond the immediate surroundings of individual UTOPÍAs, while differences in network connectivity may constrain access from other parts of the territory. These differentiated accessibility conditions provide the basis for examining the relationship between potential accessibility and observed patterns of facility use in the following section.

\subsection{Patterns of use and spatial reach of social infrastructure}
The independent location-based mobility observations reported by \textcolor{blue}{\parencite{Langle2026}} provide evidence of realized visits for 13 of the 15 UTOPÍAs. The observed facilities differ in both visitation intensity and the geographical reach of their visitors. These empirical measures are distinct from the network-based accessibility indicators: visitor counts describe activity recorded in the mobility sample, while home-to-site distances and the share of visitors residing in Iztapalapa characterize the spatial extent of observed use. The facility-level contrasts and statistical relationships are presented in Section~\ref{subsec:form_access_function}.

\begin{figure}
    \centering
    \includegraphics[width=.95\linewidth]{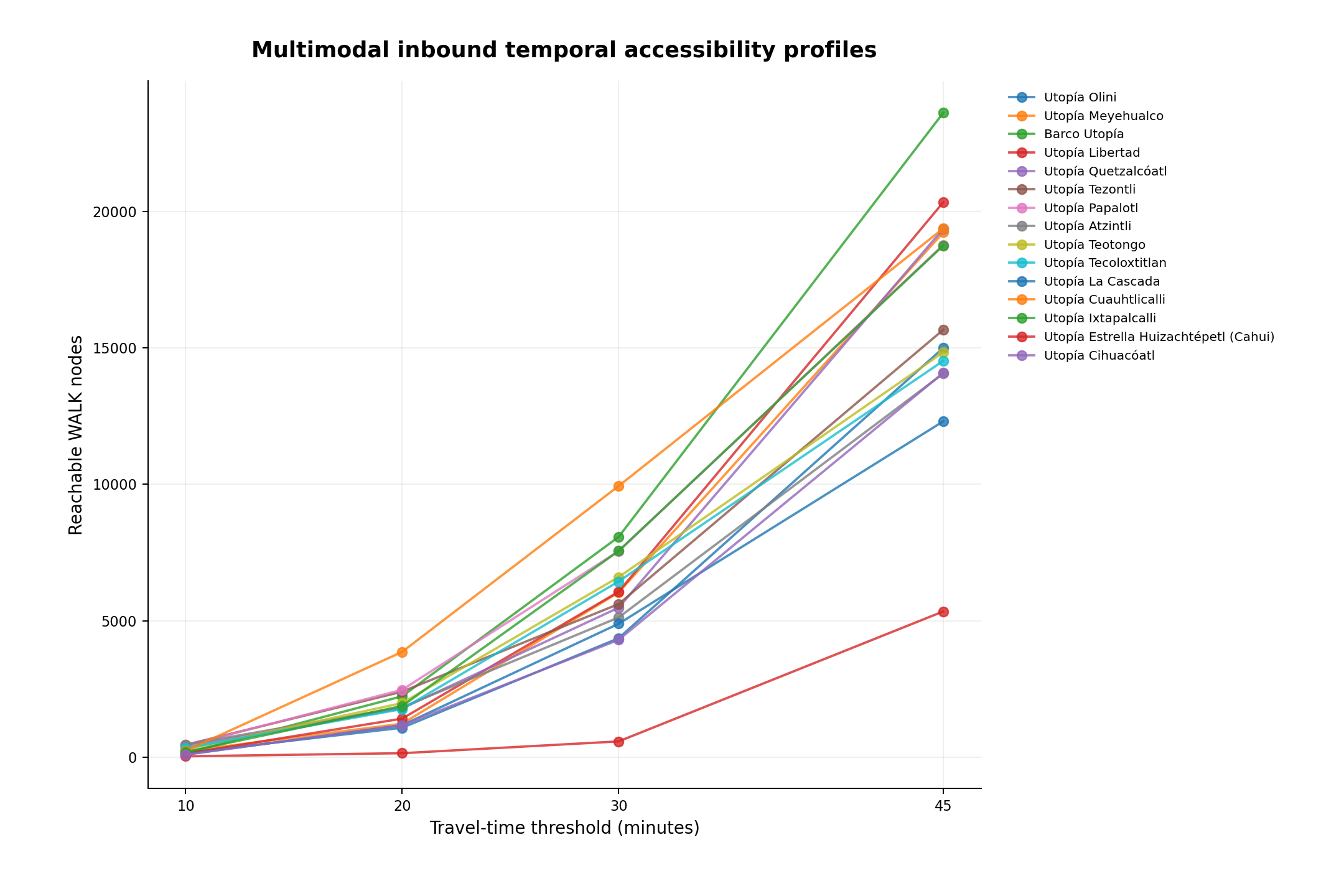}
    \caption{Multimodal inbound temporal accessibility profiles for the 15 UTOPÍAs in Iztapalapa, Mexico City. The figure compares the cumulative number of reachable pedestrian network nodes (WALK nodes) for each UTOPÍA under four travel-time thresholds (10, 20, 30, and 45 minutes). Each colored line represents an individual facility, with accessibility calculated through the integrated pedestrian and GTFS-based public transport network. The profiles illustrate how potential territorial reach expands with increasing travel-time budgets and reveal differences in the magnitude and rate of accessibility growth among facilities. Reachability is expressed as the number of accessible network nodes rather than population coverage or observed facility use.}
    \label{fig:ML_2}
\end{figure}

These temporal accessibility profiles measure potential reachability, not observed use. The independent visitation data complement these network results by describing the intensity and territorial extent of recorded visits; the two dimensions are compared below rather than treated as equivalent.

\begin{figure}
    \centering
    \includegraphics[width=1\linewidth]{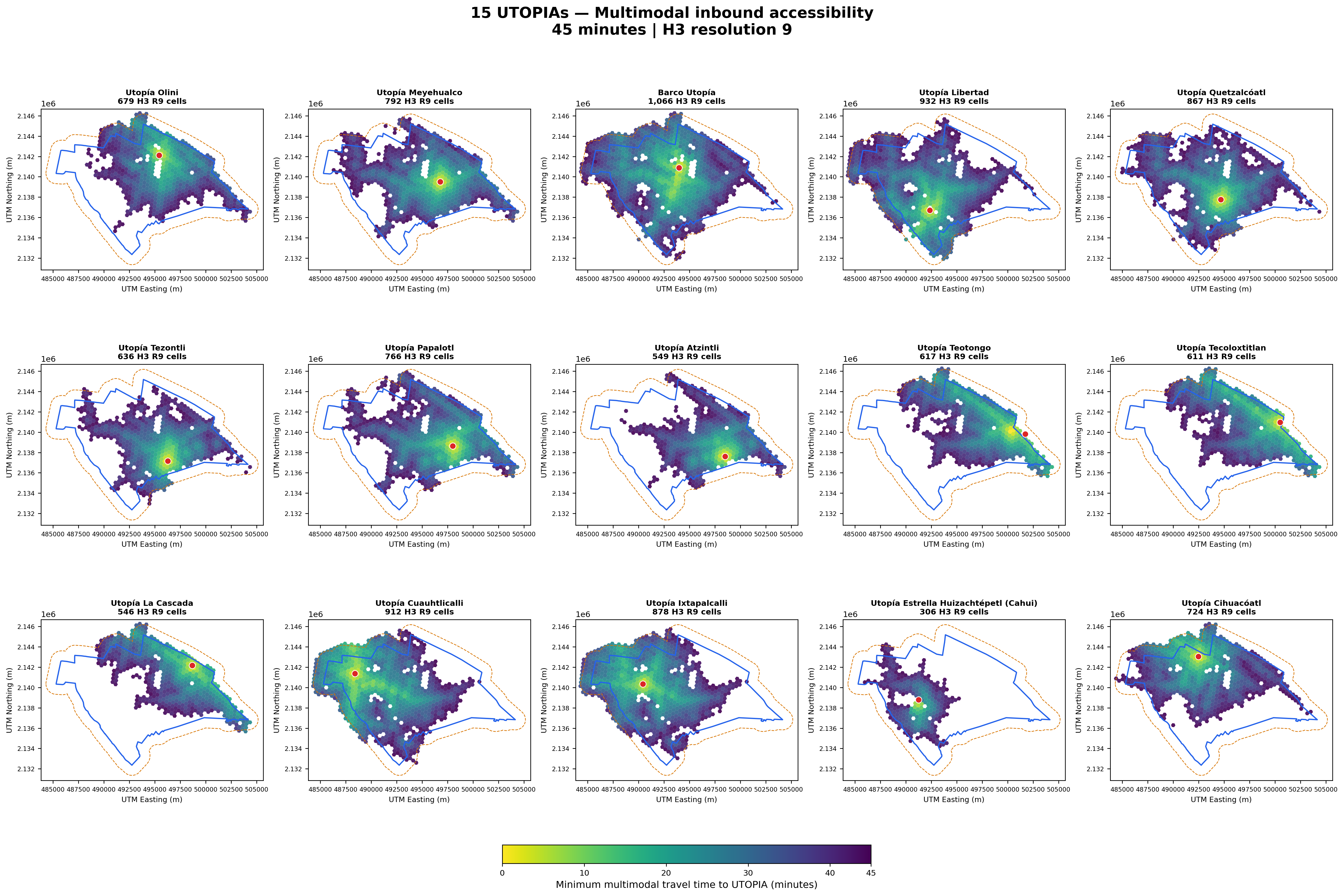}
    \caption{Spatial distribution of multimodal accessibility to the 15 UTOPÍAs in Iztapalapa, Mexico City. The maps show the modeled minimum travel time required to reach each UTOPÍA via the integrated public transport and pedestrian network, using an accessibility threshold of 45 minutes. Each panel represents a specific UTOPÍA, with colors indicating the minimum travel time (in minutes) across the modeled network. The maps illustrate spatial heterogeneity in potential accessibility across facilities, reflecting variations in location and integration into the multimodal transport system. Accessibility here represents modeled travel opportunities rather than observed visitor movements or actual facility usage. Notably, public transport infrastructure extends high-accessibility corridors (shown in green–yellow) for certain UTOPÍAs, such as Cascada and Cuauhticalli. In the case of Cuauhticalli, the intersection of two major transport systems expands the reachable area to 912 H3 cells (R9), compared to Cascada, which relies on a single transit corridor reaching 546 cells.}
    \label{fig:scatter_p}
\end{figure}

The multimodal inbound temporal accessibility profiles presented in Figure~\ref{fig:ML_2} show a substantial expansion in the number of reachable pedestrian network nodes as travel-time thresholds increase from 10 to 45 minutes. Although this expansion is observed across all 15 UTOPÍAs, its magnitude and temporal progression vary considerably among facilities. Differences become particularly pronounced at longer travel-time thresholds, indicating that the spatial reach of the integrated transport network is unevenly distributed across the study area. These findings highlight the importance of evaluating accessibility at multiple temporal scales rather than relying on a single travel-time threshold.

Building on these temporal patterns, Figure~\ref{fig:scatter_p} illustrates the spatial distribution of inbound accessibility through the integrated pedestrian and multimodal public transport network. The maps show how public transport connectivity extends the geographical reach of the UTOPÍAs beyond their immediate pedestrian catchment areas, creating corridors of accessibility that connect more distant parts of the urban territory. However, this spatial expansion is heterogeneous across facilities, reflecting differences in their location and integration with the transport network. Together, Figure~\ref{fig:ML_2} and~Figure~\ref{fig:scatter_p} provide complementary evidence of the temporal progression and spatial distribution of multimodal accessibility, highlighting the role of public transport in expanding potential interactions between populations and social infrastructure. These differentiated accessibility conditions may influence the capacity of individual UTOPÍAs to develop localized functional centralities.

\begin{figure}
    \centering
    \includegraphics[width=1\linewidth]{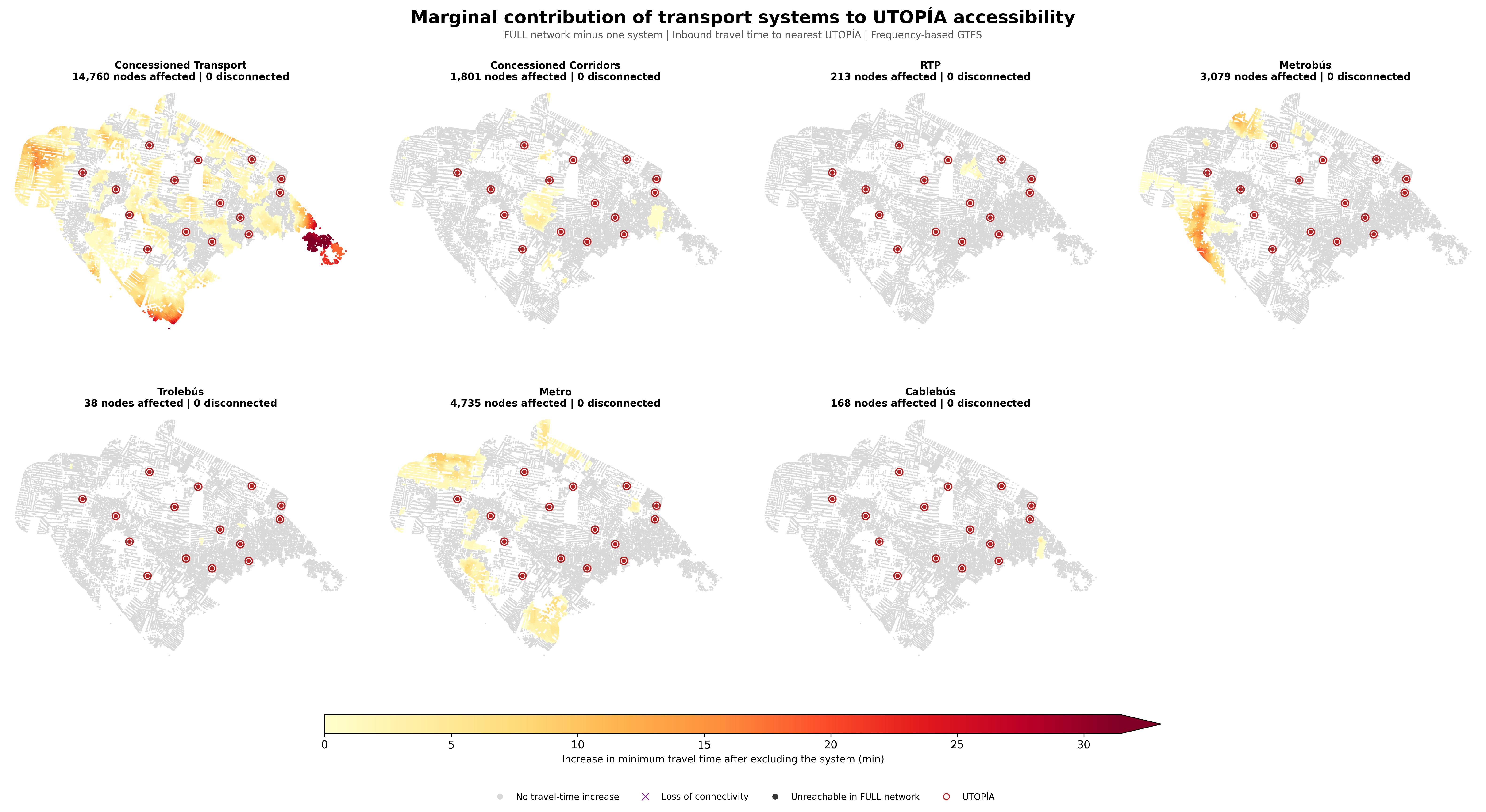}
    \caption{Marginal contribution of public transport systems to accessibility to UTOPÍA centers in Iztapalapa. Spatial distribution of changes in minimum travel time to the nearest UTOPÍA following the exclusion of individual public transport systems from the complete multimodal network. Each panel represents a leave-one-system-out scenario, with colors indicating the increase in travel time (minutes) relative to the complete network. Gray nodes indicate no increase in travel time, while red markers identify UTOPÍA locations. The number of affected pedestrian nodes is reported for each system. Across the seven exclusion scenarios, no pedestrian nodes became disconnected; however, the magnitude and spatial distribution of travel-time increases vary among transport systems. The analysis represents modeled potential accessibility rather than observed travel behavior.}
    \label{fig:map_walk}
\end{figure}

The marginal contribution analysis presented in (Figure~\ref{fig:map_walk}) provides further evidence of the differentiated role of public transport systems in shaping accessibility to UTOPÍA centers. Removing individual systems from the integrated multimodal network produces spatially heterogeneous increases in minimum travel times, even though connectivity is preserved across all exclusion scenarios. Concessioned Transport exhibits the broadest spatial effect, affecting approximately 14,760 pedestrian nodes, followed by Metrobús and Metro, whose exclusion also produces substantial accessibility changes. By contrast, the effects associated with RTP, Cablebús, and Trolebús are more spatially concentrated. These findings suggest that the contribution of individual transport systems cannot be understood solely through their physical presence or network coverage, but must also be evaluated in terms of their functional integration within the multimodal network. Together with the temporal accessibility profiles and spatial accessibility maps presented in Figure~\ref{fig:scatter_p} and Figure~\ref{fig:map_walk}, these results demonstrate how the integration of multiple transport systems shapes differentiated opportunities to access social infrastructure across Iztapalapa.

\subsection{Directional accessibility and functional centrality potential}
The combined evidence shows heterogeneous configurations of potential accessibility and realized use rather than a uniform hierarchy of functional centralities. In particular, observed visitor intensity does not increase systematically with 45-minute population coverage, while individual facilities differ in their local structural contexts and observed catchment distances (Table~\ref{tab:form_access_function}). These differences are consistent with distinct functional roles but do not, on their own, demonstrate changes in the urban centrality structure.

\begin{figure}
    \centering
    \includegraphics[width=0.85\linewidth]{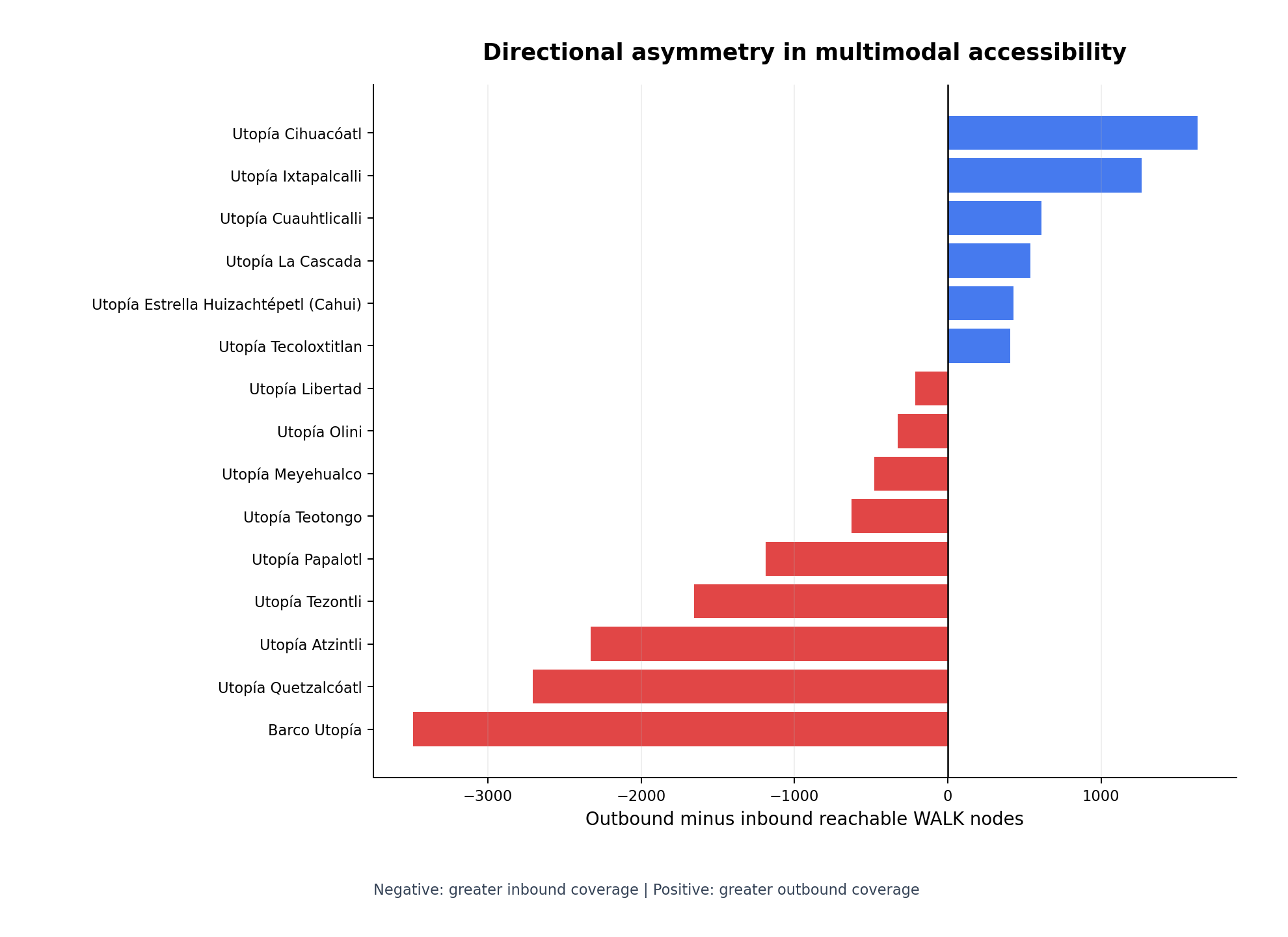}
    \caption{Directional asymmetry in multimodal accessibility to UTOPÍAs. The figure illustrates differences in the number of reachable pedestrian network nodes within a 45-minute multimodal travel-time threshold, distinguishing between inbound accessibility (nodes from which a UTOPÍA can be reached) and outbound accessibility (nodes reachable from a UTOPÍA). Asymmetry is calculated as outbound minus inbound reachable nodes; negative values indicate greater inbound coverage, whereas positive values indicate greater outbound coverage. These differences may arise from the directed structure of the transport network, route configurations, asymmetric travel times, and transfer connections. The observed asymmetries suggest that UTOPÍAs may exhibit different capacities for territorial accessibility depending on travel direction, highlighting the importance of network configuration in shaping their potential role as functional urban centralities. However, the relative contribution of each network mechanism remains to be established.
}
    \label{fig:placeholder}
\end{figure}

A further network-level distinction concerns travel direction. Figure~\ref{fig:placeholder} compares the number of pedestrian nodes reachable within 45 minutes on inbound and outbound journeys. The differences demonstrate that modeled access to a UTOPÍA need not equal the set of destinations reachable when departing from that facility. This asymmetry describes directed transport connectivity rather than observed visitor movements.

Taken together, the directional and facility-level results show why network reachability, observed visitation, and structural centrality must be evaluated separately before interpreting their relationship.

\subsection{Structural centrality and accessibility configurations}
\begin{figure}
    \centering
    \includegraphics[width=1\linewidth]{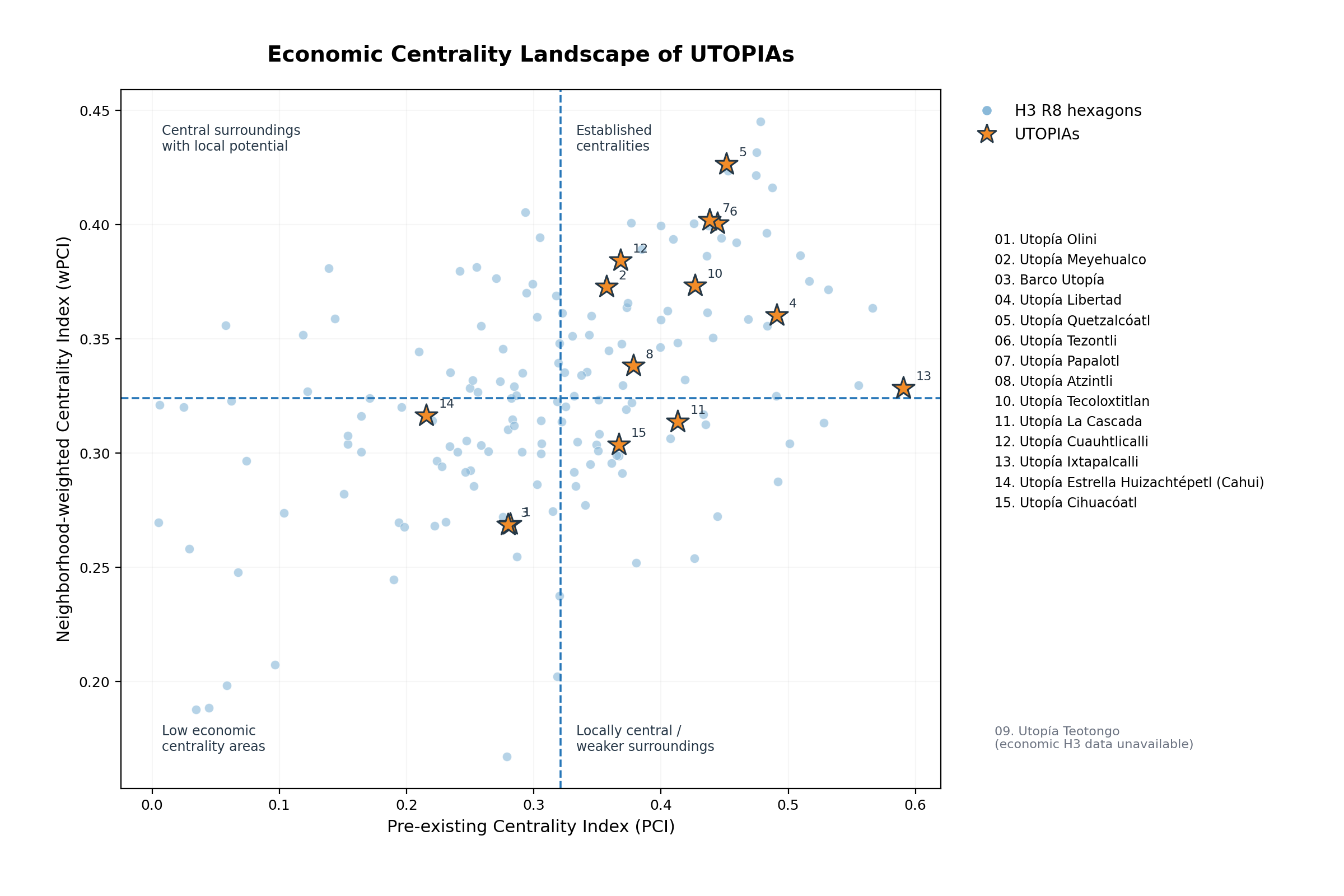}
    \caption{Economic centrality landscape of UTOPÍAs in Iztapalapa. Relationship between the local Pre-existing Centrality Index (PCI) and its spatially weighted counterpart (wPCI), calculated for H3 resolution 8 hexagons. Light-blue points represent the evaluated hexagons, while orange stars identify UTOPÍA locations, numbered according to the accompanying list. Dashed lines indicate the thresholds used to distinguish high and low centrality values, defining four combinations of local and surrounding structural conditions. These combinations differentiate areas with relatively strong local and neighboring centrality, areas with weaker local centrality but stronger surrounding conditions, and areas characterized by lower centrality in both dimensions. The figure illustrates the spatial heterogeneity of pre-existing structural centrality around UTOPÍAs, without implying observed facility use or established functional polycentricity.}
    \label{fig:HUB15_Fig02}
\end{figure}

\begin{figure}
    \centering
    \includegraphics[width=0.85\linewidth]{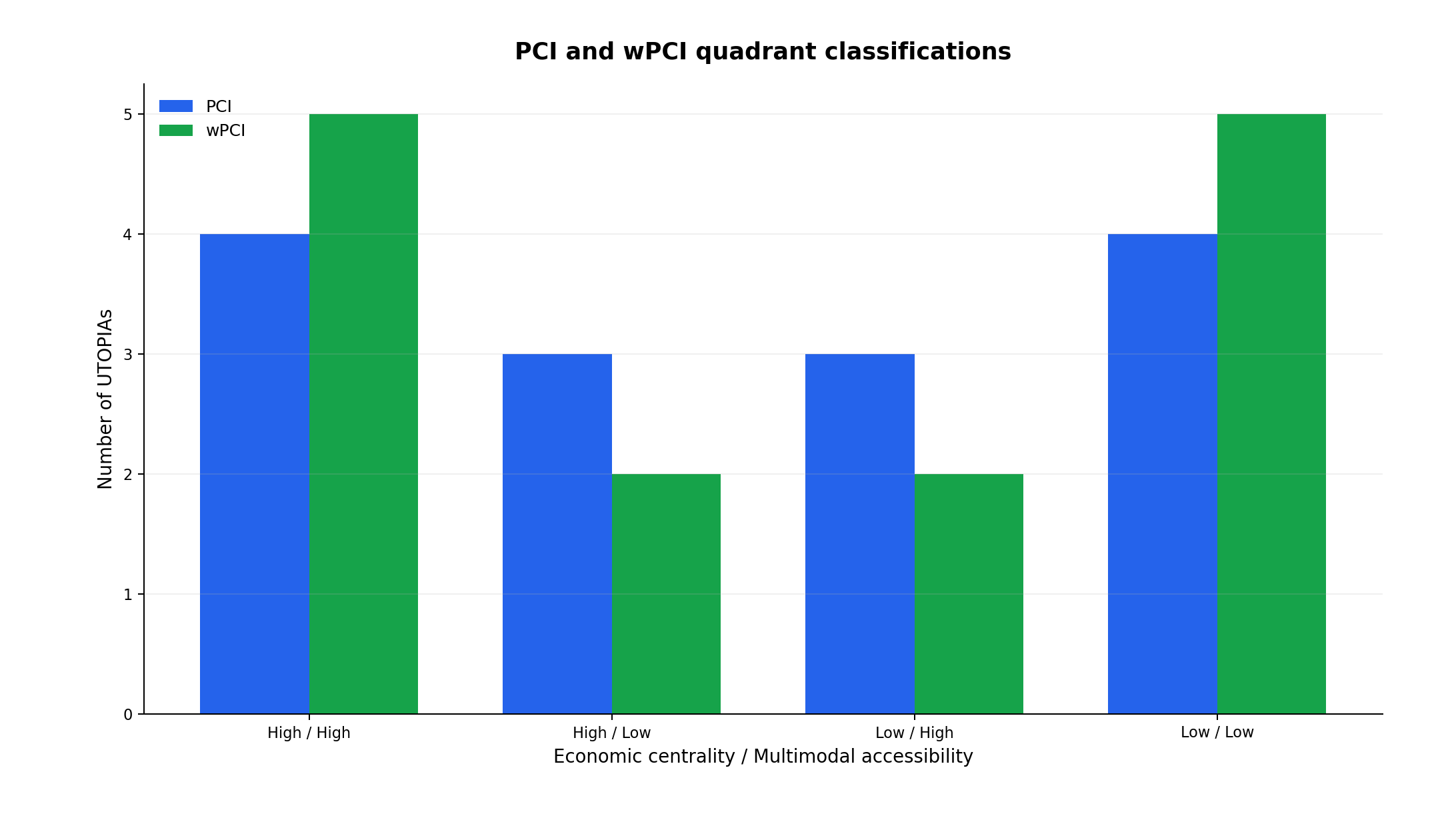}
    \caption{Comparison of PCI- and wPCI-based quadrant classifications of the 15 UTOPÍAs in Iztapalapa. Distribution of UTOPÍAs across four quadrants defined by the combination of high or low pre-existing structural centrality and high or low multimodal accessibility. Blue bars represent classifications based on the local Pre-existing Centrality Index (PCI), while green bars represent classifications using its spatially weighted counterpart (wPCI). The quadrants distinguish facilities with high centrality and high accessibility (High/High), high centrality and low accessibility (High/Low), low centrality and high accessibility (Low/High), and low centrality and low accessibility (Low/Low). Differences between PCI and wPCI classifications illustrate how incorporating neighboring economic conditions modifies the spatial characterization of potential functional centralities.}
    \label{fig:HUB13_Fig03}
\end{figure}

The comparison of structural centrality and multimodal accessibility reveals different combinations of local urban conditions across the 15 UTOPÍAs. Figures~\ref{fig:HUB15_Fig02} and~\ref{fig:HUB13_Fig03} describe these configurations using PCI and its neighborhood-weighted counterpart, wPCI. These are structural--accessibility quadrants; they should not be interpreted as direct classifications of observed use.

The independent visitation data provide a separate test of accessibility--use alignment. As shown in Table~\ref{tab:form_access_function}, 45-minute population coverage has no clear association with observed visitation intensity, indicating that modeled access and recorded use cannot be treated as interchangeable.

Figures \ref{fig:HUB15_Fig02} and  \ref{fig:HUB13_Fig03} provide complementary perspectives on the relationship between pre-existing structural centrality and multimodal accessibility in Iztapalapa. Figure~\ref{fig:HUB15_Fig02} illustrates the spatial heterogeneity of local (PCI) and neighborhood-weighted (wPCI) economic centrality, while Figure~\ref{fig:HUB13_Fig03} examines how these alternative measures modify the classification of the 15 UTOPÍAs across centrality–accessibility quadrants. The differences between PCI- and wPCI-based classifications suggest that the surrounding urban economic context contributes to the characterization of potential functional centralities, beyond the conditions observed at individual facility locations.

The differences between PCI- and wPCI-based quadrants demonstrate the sensitivity of structural--accessibility classification to the surrounding urban context. Because these quadrants do not directly incorporate visitation, their functional interpretation requires the complementary observed-use evidence presented later in this section.

\subsection{Pedestrian proximity and territorial reach}
The observed contrasts indicate that proximity has more than one empirical dimension. Network-based pedestrian reachability measures the opportunity to reach a facility without boarding public transport, whereas the median distance traveled by observed visitors describes the territorial extent of recorded use. These indicators do not measure the same process and need not vary together.

In the available sample, walk-only accessibility share exhibits a positive rank association with observed visitation, although the bootstrap uncertainty is substantial (Table~\ref{tab:form_access_function}). At the same time, facilities such as Olini and Meyehualco draw observed visitors from comparatively broad catchments, whereas Tezontli shows a more localized catchment. These differences support a differentiated account of proximity and territorial reach, without establishing that new urban centralities have already formed.

\subsection{Pre-existing urban structure, accessibility, and realized function}
\label{subsec:form_access_function}

To examine whether the spatial conditions identified through the structural and multimodal analyses correspond to realized urban function, the indicators of pre-existing centrality and potential accessibility were contrasted with independent evidence of observed visitation. Realized-use indicators were
obtained from the location-based mobility analysis reported by \textcolor{blue}{\parencite{Langle2026}}, which provides comparable observations for 13 of the 15 UTOPÍAs included in our network. These data were used as an independent empirical characterization of realized function rather than as an alternative measure of accessibility. The spatial and accessibility analyses cover all 15 UTOPÍAs, including Teotongo. Observed visitation is available for 13 facilities.

The joint comparison reveals that pre-existing urban structure, potential accessibility, and realized use capture related but non-equivalent dimensions of functional centrality. In particular, potential multimodal accessibility did not show a direct correspondence with visitation intensity. Population coverage within 45 minutes was weakly associated with log-transformed observed visitors (Pearson $r=-0.273$; Spearman $\rho=-0.294$), and neither association was supported by the permutation tests ($p_{\mathrm{perm}}=0.372$ and
$p_{\mathrm{perm}}=0.327$, respectively). Population-weighted median travel time showed similarly modest associations with visitation ($r=0.316$, $p_{\mathrm{perm}}=0.286$; $\rho=0.347$, $p_{\mathrm{perm}}=0.246$).
Thus, within the facilities for which observed-use data are available, greater population coverage or shorter multimodal travel times did not systematically correspond to higher visitation intensity.

A clearer relationship emerged between pre-existing urban structure and the spatial scale of realized function. PCI was positively associated with the share of visitors originating within Iztapalapa ($r=0.656$, $p_{\mathrm{perm}}=0.023$; $\rho=0.594$, $p_{\mathrm{perm}}=0.045$).
The bootstrap 95\% confidence interval for the Pearson coefficient remained positive ($0.301$--$0.857$), while leave-one-out estimates were consistently positive for both Pearson ($0.551$--$0.701$) and Spearman
($0.473$--$0.682$) correlations. This relationship therefore remained stable when individual UTOPÍAs were sequentially excluded from the analysis, indicating that it was not driven by a single facility.

\begin{figure}
    \centering
    \includegraphics[width=1\linewidth]{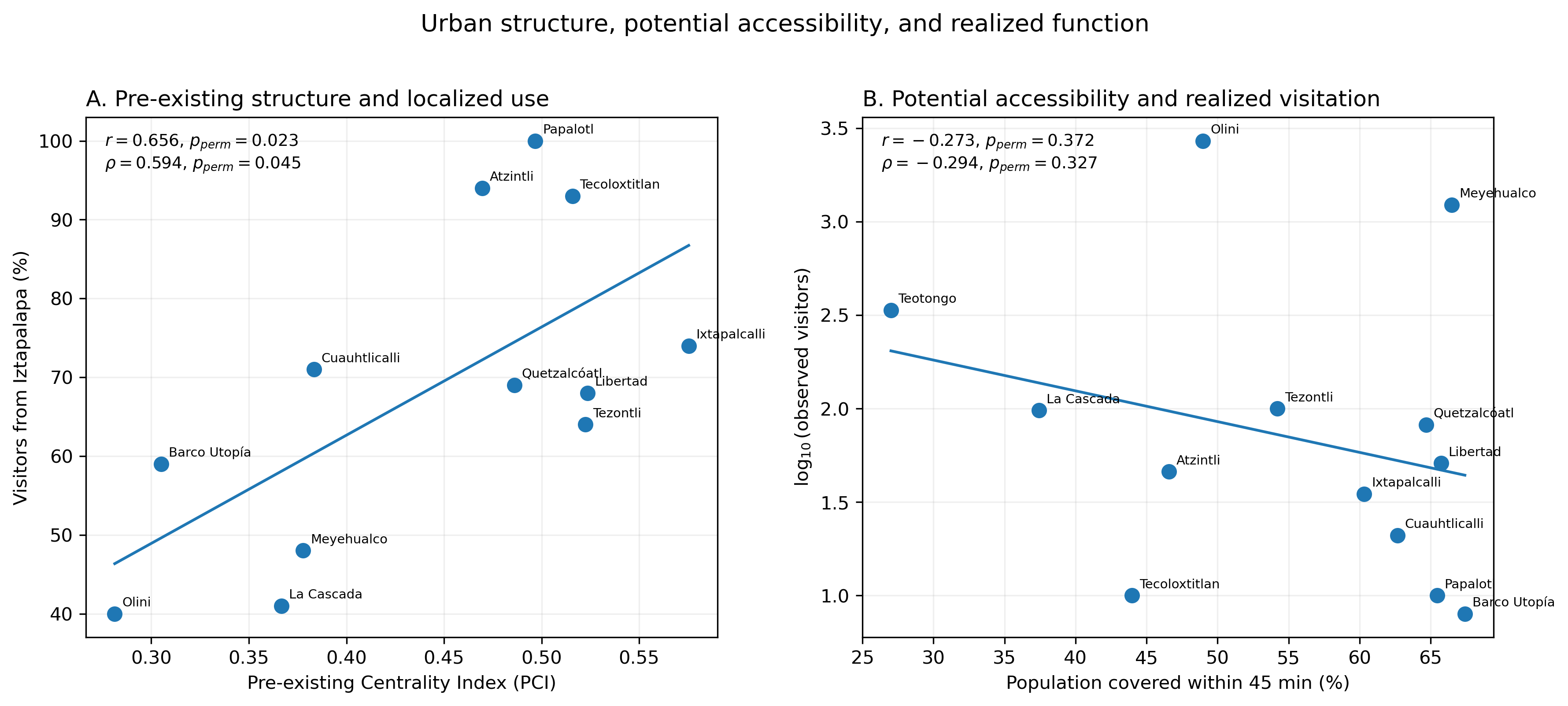}
    \caption{Relationship between urban structure, potential accessibility, and realized function across the UTOPÍAs.  \textbf{(A)} Pre-existing structural centrality (PCI) and the share of observed visitors originating within Iztapalapa ($n=12$). \textbf{(B)} Population coverage within 45 minutes and log-transformed observed visitation ($n=13$). Panel A shows a positive association supported by permutation and sensitivity analyses between pre-existing structural centrality and localized realized use, whereas Panel B illustrates the absence of a simple correspondence between potential multimodal accessibility and visitation intensity. Observed visitation indicators are derived from the independent location-based mobility dataset reported by \textcolor{blue}{\parencite{Langle2026}}. Lines indicate linear trends for visualization only.}
    \label{fig:Accesibility_Function}
\end{figure}

Figure~\ref{fig:Accesibility_Function} provides a visual comparison of these contrasting relationships. Panel A shows a positive association between pre-existing structural centrality (PCI) and the proportion of visitors originating within Iztapalapa, suggesting that stronger local economic contexts are associated with more geographically localized patterns of use. In contrast, Panel B shows no clear association between 45-minute multimodal population coverage and observed visitation intensity. Together, these findings indicate that pre-existing urban structure and potential accessibility capture distinct dimensions of the relationship between social infrastructure and realized urban function.

\begin{table*}[htbp]
\centering
\caption{Selected associations between pre-existing urban structure, potential accessibility, and realized function.}
\label{tab:form_access_function}

\small
\setlength{\tabcolsep}{4pt}
\renewcommand{\arraystretch}{1.15}

\begin{tabularx}{\textwidth}{
    >{\raggedright\arraybackslash}p{2.2cm}
    >{\raggedright\arraybackslash}X
    >{\centering\arraybackslash}p{0.55cm}
    >{\centering\arraybackslash}p{1.05cm}
    >{\centering\arraybackslash}p{1.05cm}
    >{\centering\arraybackslash}p{1.15cm}
    >{\centering\arraybackslash}p{1.05cm}
    >{\raggedright\arraybackslash}p{2.4cm}
}

\toprule

\textbf{Dimension} &
\textbf{Relationship} &
\textbf{$n$} &
\textbf{Pearson $r$} &
\textbf{$p_{\mathrm{perm}}$} &
\textbf{Spearman $\rho$} &
\textbf{$p_{\mathrm{perm}}$} &
\textbf{Interpretation} \\

\midrule

Structure--function &
PCI $\leftrightarrow$ visitors from Iztapalapa (\%) &
12 &
0.656 &
0.023 &
0.594 &
0.045 &
Consistent positive association \\

Structure--function &
WPCI $\leftrightarrow$ median home-to-site distance &
12 &
-0.538 &
0.082 &
-0.528 &
0.090 &
Suggestive, directionally robust \\

\midrule

Accessibility--function &
45-min population coverage $\leftrightarrow$ log$_{10}$ visitors &
13 &
-0.273 &
0.372 &
-0.294 &
0.327 &
Weak / no clear association \\

Accessibility--function &
Weighted median travel time $\leftrightarrow$ log$_{10}$ visitors &
13 &
0.316 &
0.286 &
0.347 &
0.246 &
Weak / no clear association \\

Accessibility--function &
Walk-only share $\leftrightarrow$ log$_{10}$ visitors &
13 &
0.245 &
0.430 &
0.591 &
0.037 &
Exploratory monotonic association \\

\midrule

Structure--accessibility &
WPCI $\leftrightarrow$ 90th-percentile travel time &
14 &
-0.434 &
0.124 &
-0.490 &
0.074 &
Suggestive, directionally robust \\

\bottomrule

\end{tabularx}

\vspace{0.5em}

\begin{minipage}{\textwidth}
\footnotesize
\textit{Note:} PCI denotes the Pre-existing Centrality Index and WPCI its distance-weighted neighborhood formulation. Realized-use indicators are derived from the independent location-based mobility dataset reported by \textcolor{blue}{\parencite{Langle2026}}. Visitor counts correspond to unique visitors observed in the LBS sample and are not official attendance counts. Two-sided permutation tests were based on 10,000 permutations. The reported $n$ values and coefficients reproduce the available statistical output and require verification against the final 15-facility PCI/WPCI dataset, including Teotongo. Visitation data are available for 13 facilities; a smaller structural--use sample cannot be attributed to Teotongo being geographically outside the study area. Comparisons between accessibility and realized use use $n=13$. Interpretations reflect coefficient magnitude, permutation tests, bootstrap uncertainty, and leave-one-out stability and should not be interpreted as evidence of causality.
\end{minipage}

\end{table*}

A complementary pattern was observed for neighborhood-weighted structural centrality. WPCI was negatively associated with the median home-to-site distance of observed visitors ($r=-0.538$, $p_{\mathrm{perm}}=0.082$;
$\rho=-0.528$, $p_{\mathrm{perm}}=0.090$).
Although these permutation probabilities do not provide strong evidence at conventional significance thresholds, the direction and magnitude of the relationship were consistent across Pearson and Spearman estimates.
Furthermore, leave-one-out coefficients remained negative in every case (Pearson: $-0.631$ to $-0.417$;
Spearman: $-0.642$ to $-0.431$).
Taken together with the PCI results, these patterns indicate that pre-existing structural centrality is more consistently associated with the territorial scale of realized use than with the absolute intensity of visitation: facilities embedded in stronger structural contexts tend to exhibit more spatially localized catchments within Iztapalapa.

Pre-existing structural centrality and potential multimodal accessibility were also only partially aligned. WPCI showed a moderate negative association with the population-weighted 90th-percentile travel time ($r=-0.434$, $p_{\mathrm{perm}}=0.124$; $\rho=-0.490$, $p_{\mathrm{perm}}=0.074$).
Although uncertainty remains substantial, the relationship was directionally stable under leave-one-out analysis, with Pearson coefficients ranging from $-0.617$ to $-0.349$ and Spearman coefficients from $-0.753$ to $-0.379$. This pattern suggests that the broader structural context surrounding a UTOPÍA
may be related to lower travel-time penalties for the less-accessible parts of the population, although the small number of facilities warrants a cautious interpretation.

Pedestrian-scale accessibility displayed a different relationship with realized use. The share of the represented population able to reach a UTOPÍA through walk-only routes was weakly related to log-transformed visitation under a linear specification ($r=0.245$, $p_{\mathrm{perm}}=0.430$), but showed a stronger rank-based association ($\rho=0.591$, $p_{\mathrm{perm}}=0.037$).
The bootstrap interval for the Spearman coefficient was wide and marginally included zero ($-0.009$--$0.836$), but the positive association remained stable in the leave-one-out analysis
($\rho=0.524$--$0.676$).

The divergence between the Pearson and Spearman estimates suggests a possible monotonic but non-linear relationship between pedestrian-scale proximity and observed use. Given the small sample, this result is treated as exploratory rather than as evidence of a systematic effect.

Facility-level contrasts further illustrate why potential accessibility and realized function should not be treated as interchangeable. Olini combines relatively low pre-existing structural centrality (PCI $=0.281$; WPCI $=0.285$) and intermediate 45-minute population coverage (49.0\%) with the highest observed visitor count in the independent mobility sample (2,709 visitors) and a median home-to-site distance of 7.5~km.
Meyehualco similarly exhibits a broad realized catchment, with 1,231 observed visitors and a median home-to-site distance of 6.7~km. In contrast, Tezontli is embedded in a comparatively strong structural context (PCI $=0.522$; WPCI $=0.440$) and reaches 54.2\% of the represented population within 45 minutes, but its observed catchment is considerably more localized, with a median home-to-site distance of 1.1~km. These contrasts show that visitation intensity, territorial reach, and localized neighborhood use represent different expressions of realized function rather than interchangeable measures of functional centrality.

Overall, the results do not support a simple sequence in which greater pre-existing centrality produces greater accessibility and greater accessibility in turn produces higher visitation. Instead, they reveal a differentiated relationship between urban form, accessibility, and realized function.
Pre-existing structural centrality is more consistently associated with the \emph{spatial scale} of realized use, particularly the localization of catchments within Iztapalapa, whereas potential multimodal accessibility alone does not explain visitation intensity. At the same time, the exploratory relationship between walk-only accessibility and observed use suggests that proximity may constitute an additional pathway through which localized functional roles develop. These heterogeneous configurations provide the
empirical basis for subsequently evaluating whether individual UTOPÍAs reinforce pre-existing centralities, exhibit evidence consistent with emerging functional centralities, or remain functionally constrained.

\clearpage

\section{Discussion}
\label{sec:discussion}

\subsection{Multimodal accessibility and functional proximity}
\label{sec:discussion-proximity}

The baseline, mode-exclusion, and directional scenarios provide complementary evidence that proximity to social infrastructure is a property of the operating mobility network rather than a simple function of geographic distance. The baseline results in the accessibility results show that travel to the UTOP\'{I}As combines in-vehicle movement with substantial walking and waiting costs (Table~\ref{tab:travel_time_components}). Thus, the location of a facility relative to a transport corridor is insufficient to characterize its accessibility: pedestrian connections, service frequency, and transfers jointly determine the time required to reach it. This distinction matters for interpreting social infrastructure as part of an interconnected urban socioecological system rather than as a collection of isolated destinations.

The mode-exclusion results further qualify this interpretation. Removing individual transport systems changes modeled accessibility unevenly across the pedestrian network, with concessioned services producing a particularly extensive spatial effect. This is evidence of differentiated \emph{network contributions}, not a direct measure of observed ridership, operational reliability, or system-wide resilience. The findings indicate that the potential reach of UTOP\'{I}As depends on connections among services with different spatial roles; consequently, an assessment restricted to high-capacity or institutionally operated modes would overlook relevant parts of the accessibility structure. Similarly, the inbound--outbound comparison in the results on functional centralities shows that reachability is direction-dependent. A facility may be relatively accessible as a destination without providing an equivalent set of opportunities for onward travel. These asymmetries should not be interpreted as observed trip patterns, but they demonstrate why a single undirected proximity measure can obscure constraints relevant to everyday access.

Together, these scenarios support a relational interpretation of \emph{Potential Function}: the opportunities for reaching social infrastructure generated by the combined pedestrian and public transport system. They also delimit that concept. Modeled reachability under the specified service and departure assumptions does not establish affordability, safety, comfort, reliability, or equitable access. The contribution of the multilayer approach is to identify where the transport network enables or constrains potential interaction, providing a basis for examining whether those opportunities correspond to realized use.

\subsection{From potential accessibility to realized urban function}
\label{sec:discussion-function}

The comparison with independent visitation observations shows why Potential Function and \emph{Realized Function} must remain analytically distinct. As reported in the results on observed use and the form--function relationships, population coverage within 45 minutes does not exhibit a clear association with observed visitation intensity (Figure~\ref{fig:Accesibility_Function}). A larger modeled catchment therefore cannot be assumed to generate more visitors. Conversely, facilities with modest structural centrality or intermediate accessibility can attract substantial observed use. The contrasting profiles of Olini, Meyehualco, and Tezontli illustrate that visitor volume, geographic catchment, and local orientation describe different aspects of realized function.

The positive association reported between pre-existing PCI and the proportion of visitors originating within Iztapalapa suggests a different relationship: facilities embedded in stronger local structural contexts may serve more geographically localized visitor populations. This is an association, not evidence that pre-existing economic activity causes localized use. Moreover, the structural--use estimates currently reported for a subset of facilities require confirmation against the final PCI/wPCI dataset covering all 15 UTOP\'{I}As. The small visitation sample, the two facilities without observed-use information, and the observational nature of the mobility data warrant caution in interpreting both significant and non-significant coefficients. In particular, the absence of a clear coverage--visitation association does not establish that accessibility is unimportant; it indicates that accessibility alone does not explain the observed variation in use.

This separation between potential and realized function is central to the form--function argument. Pre-existing urban structure describes the territorial context of opportunities; the GTFS-HUB network estimates the accessibility those locations can offer; and the independent visitation data provide a partial observation of actual interaction with social infrastructure \textcolor{blue}{\parencite{Langle2026}}. None of these dimensions substitutes for the others. Their comparison identifies mismatches and complementary configurations that would remain hidden if functional centrality were inferred solely from transport connectivity, facility provision, or visitor counts. In this sense, the framework contributes to systems-oriented accounts of cities in which urban properties depend on interactions among spatial, infrastructural, and social components \textcolor{blue}{\parencite{Batty2005, Frank2017, Cioffi2016}}.

\subsection{Functional centrality, localized polycentricity, and sustainability}
\label{sec:discussion-polycentricity}

The combined results offer evidence of heterogeneous \emph{functional roles} among the UTOP\'{I}As, but they do not establish that the facilities have reorganized the wider urban system into a consolidated polycentric structure. Differences in PCI/wPCI, accessibility, visitor intensity, and catchment orientation suggest that some facilities operate within reinforced pre-existing contexts while others may represent emerging sites of social interaction. These categories are useful for interpreting local configurations; they should not be treated as proof of a temporal transition, because the analysis does not compare the urban system before and after the establishment of each facility.

This distinction clarifies the meaning of \emph{Emerging Functional Centrality} in the present study. It is not equivalent to high network centrality, the mere presence of social infrastructure, or a large modeled population catchment. Rather, it denotes a plausible localized concentration of function that must be assessed through the relationship between structural context, opportunities for access, and observed use. The available evidence can characterize that relationship at the facility level, but cannot yet demonstrate sustained growth in interactions, redistribution of activities from established centers, or changes in metropolitan morphology. Localized functional differentiation is therefore a more defensible empirical interpretation than an asserted transformation to polycentricity. This distinction is consistent with the need to separate morphological and functional dimensions in the study of urban centers \textcolor{blue}{\parencite{Batty2005, Burger2014, Cioffi2016}}.

The sustainability implications follow from this distinction rather than from an assumed causal effect of the UTOP\'{I}As. Social infrastructure may broaden the spatial distribution of opportunities, but the capacity to use those opportunities remains conditioned by multimodal travel times, pedestrian connections, service availability, and the surrounding urban context. The modeled contributions of different transport systems and the directional asymmetries indicate that improving access cannot be reduced to adding facilities or optimizing a single transport mode. Likewise, observed visitation does not by itself demonstrate equitable participation: the available data do not identify all potential users, unmet demand, barriers faced by non-visitors, or changes in social well-being.

The prevalence of Mexico City’s concession-based public transport system—often classified as 'informal' \textcolor{blue}{\parencite{Cervero2007}}—mirrors the vehicle profiles typical of informal transit across developing nations. Furthermore, the nine core characteristics of public transport defined by \textcolor{blue}{\parencite{Olowosegun2021}}—accessibility, fares, travel time, waiting time, passenger comfort, driver attitude, safety, speed, and vehicle amenities—further highlight the system's marked degree of informality. This dynamic generates both costs and benefits, shaping equity outcomes as well as the well-being of users and the urban environment. As presented in Table 4, our findings reveal that accessibility to the Utopias relies heavily on this concession-based system. Consequently, comprehensive accessibility evaluations must incorporate this transit mode rather than relying solely on travel time or distance metrics.

For planning, the results support evaluating investments in social infrastructure jointly with the mobility networks that connect residents to them. Facility-level differences in accessibility and realized use can help identify where further investigation of service provision, transfers, pedestrian conditions, or local demand is warranted. They do not, on their own, establish which intervention would increase attendance or produce environmental benefits. The broader contribution to Sustainability Science is thus an empirically bounded account of how urban opportunities and their realization can diverge, with implications for proximity-oriented planning and mobility justice rather than a direct measurement of sustainability outcomes.

\subsection{Methodological contributions, limitations, and future research}
\label{sec:discussion-limitations}

Methodologically, the study integrates four complementary analytical perspectives: baseline multimodal travel times and population coverage, mode-exclusion comparisons, directional accessibility, and the joint evaluation of pre-existing structure and observed use. The GTFS-HUB representation combines pedestrian links, route-specific services, waiting, in-vehicle travel, and transfers, while the PCI/wPCI indicators and independent visitation observations place modeled accessibility in its urban and social context. The main contribution is not a new single measure of centrality, but a reproducible analytical distinction between structural context, Potential Function, and Realized Function. This distinction allows claims about connectivity to be tested against, rather than conflated with, observed activity.

Several limitations constrain inference. Accessibility is modeled for a specified weekday and departure time using available GTFS service information; it does not represent the full temporal variability of operations, congestion, irregular service, affordability, perceived safety, or individual travel preferences. Mode exclusion quantifies changes in modeled reachability under a counterfactual network configuration, not actual disruption responses or resilience. Inbound and outbound accessibility describe network directionality, not observed flows. Population-weighted origins and spatial aggregation introduce scale and representation choices, while the single-district case limits direct generalization to other urban settings.

Realized Function is represented by an independent, aggregated location-based visitation sample, not official attendance records or a census of users. Its coverage is incomplete across the 15 facilities and may reflect sampling and representativeness limitations. Cross-sectional associations, particularly those based on small samples, cannot determine causality or establish the emergence of centralities over time. The final structural--use coefficients and sample sizes should also be reconciled with the corrected PCI/wPCI coverage for Teotongo before publication. These qualifications are essential to avoid interpreting an analytical typology as evidence of a completed urban transition.

Future work should combine time-varying accessibility with longitudinal and more representative measures of facility use, including information about non-users and unmet demand. Repeated observations could test whether changes in connectivity precede changes in visitation, whether localized interactions persist, and whether activity becomes redistributed among urban centers. Comparative applications to other forms of social infrastructure and metropolitan settings could then assess how transferable the proposed form--accessibility--function relationships are. Until such evidence is available, the present findings establish a framework for evaluating potential and realized urban functions, while leaving the long-term emergence of structural polycentricity as an open empirical question.

\section{Conclusion}
\label{sec:conclusion}

This study developed a multilayer analytical framework to examine the relationships between pre-existing urban structure, multimodal accessibility, and the realized function of social infrastructure in Iztapalapa, Mexico City. By integrating pedestrian connectivity, public transport networks, spatial indicators of urban centrality, and observed visitation patterns, the analysis provides a relational perspective on the conditions under which social infrastructure may contribute to emerging functional centralities within an urban socioecological system.

The findings reveal substantial differences among the 15 UTOPÍAs in terms of their structural context, multimodal accessibility, and observed use. The baseline accessibility analysis, mode-exclusion experiments, and directional comparisons demonstrate that functional proximity depends not only on the spatial distribution of facilities but also on the configuration and operation of mobility networks. Moreover, the relationships between potential accessibility and observed visitation indicate that connectivity alone does not adequately explain the realized function of social infrastructure. These results highlight the importance of distinguishing the availability of urban opportunities from the conditions that enable their effective use.

From a mobility justice and urban sustainability perspective, the findings suggest that distributing social infrastructure across underserved territories is a necessary but insufficient strategy for improving equitable access to urban opportunities. The potential contribution of UTOPÍAs to proximity-oriented urban development depends on their integration with reliable, interconnected, and territorially inclusive mobility systems. Nevertheless, the present analysis does not directly establish changes in mobility equity, environmental performance, or long-term urban restructuring; these remain important questions for further investigation.

The proposed framework contributes to urban sustainability research by linking structural centrality, potential accessibility, and realized urban function within a common analytical approach. Future research should incorporate longitudinal visitation records, more detailed origin--destination observations, temporal variations in transport services, and evidence of social appropriation to evaluate how these relationships evolve. Additional methodological limitations and opportunities for refinement are discussed in Appendix~A.5.

Ultimately, the integration of social infrastructure into multimodal mobility networks does not, by itself, demonstrate the emergence of a structurally polycentric urban system. Although the analysis identifies differentiated patterns of pre-existing centrality, potential accessibility, and observed use, these dimensions do not necessarily converge into a unified process of urban restructuring. Rather than establishing a completed transition toward polycentricity, the findings identify functional relationships and conditions through which localized centralities may emerge. This distinction is fundamental for evaluating the contribution of social infrastructure to more accessible, equitable, and socioecologically sustainable urban systems.

\section*{Acknowledgements}
\begin{itemize}
    \item All participants provided prior informed consent for the use of the materials generated during these activities in academic research.
    \item The authors thank Where Is My Transport (WIMT) for providing access to the GTFS CdMx Public Transport database from their RUMBO App.
    \item The authors acknowledge the use of AI-based language assistance tools to support the grammar and orthography of the manuscript. All interpretations, analyses, and conclusions remain the responsibility of the authors.
 \end{itemize}
 
\printbibliography

\clearpage

\appendix

\section{Mathematical Formulation of Accessibility, Spatial Friction, and Multilayer Connectivity}

\subsection{GTFS-Based Proximity and Accessibility Calculations}

Let \(U=\{u_1,\dots,u_m\}\) be the set of UTOPÍAS and \(S=\{s_1,\dots,s_n\}\) the set of public transport stops and stations derived from the GTFS-based transport layers.

For each UTOPÍA \(u_i\), spatial proximity to the transport system was calculated by identifying the nearest transport node:

\begin{equation}
d(u_i,s_j)=\min_{s_j \in S} \; \text{dist}(u_i,s_j)
\end{equation}

where \(\text{dist}(u_i,s_j)\) represents the spatial distance between the UTOPÍA and the nearest public transport stop or station.

Assuming a constant pedestrian speed, the complementary walking time required to access the transport system is:

\begin{equation}
t^{walk}_{ij}=\frac{d(u_i,s_j)}{v_{walk}}
\end{equation}

where \(v_{walk}\) is the assumed walking speed.

For the GTFS-based public transport network, each route segment between consecutive stops \(s_i\) and \(s_j\) was weighted according to scheduled travel time:

\begin{equation}
t^{pt}_{ij}=a_j-d_i
\end{equation}

where:

\begin{itemize}
    \item \(d_i\) is the scheduled departure time from stop \(s_i\),
    \item \(a_j\) is the scheduled arrival time at stop \(s_j\).
\end{itemize}

These temporal weights were derived directly from GTFS operational schedules, including stop sequences and scheduled arrival/departure times. Consequently, accessibility reflects not only geometric separation, but also the operational structure of the multimodal transport system.

The total multimodal travel time between an origin node \(o\) and a UTOPÍA \(u\) is defined as the minimum accumulated cost over all feasible paths \(P_{ou}\):

\begin{equation}
T(o,u)=\min_{p \in P_{ou}} \sum_{(i,j)\in p} w_{ij}
\end{equation}

where \(w_{ij}\) corresponds to the edge weight expressed in minutes and may represent:

\begin{itemize}
    \item walking time,
    \item in-vehicle travel time,
    \item or intermodal transfer time.
\end{itemize}

Accessibility under a maximum travel-time threshold \(\tau\) is defined as:

\begin{equation}
A(o)=
\begin{cases}
1, & \text{if } \min_{u \in U} T(o,u) \leq \tau \\
0, & \text{otherwise}
\end{cases}
\end{equation}

In this study, a temporal threshold of:

\begin{equation}
\tau = 45 \text{ minutes}
\end{equation}

was used to identify areas functionally connected to at least one UTOPÍA.

A continuous accessibility measure is also defined as:

\begin{equation}
A_c(o)=\min_{u \in U} T(o,u)
\end{equation}

where lower values indicate greater effective proximity to the UTOPÍAS network.

\subsection{Spatial Friction and Operational Accessibility}

Within the multilayer mobility system, accessibility is conditioned by spatial friction, understood as the cumulative impedance associated with movement across the urban network.

For each edge \(e_{ij}\), the generalized friction cost is represented by:

\begin{equation}
f_{ij}=t^{GTFS}_{ij}+\phi_{ij}
\end{equation}

where:

\begin{itemize}
    \item \(t^{GTFS}_{ij}\) is the travel-time cost derived from GTFS operational schedules,
    \item \(\phi_{ij}\) is an additional impedance term associated with transfers, modal discontinuities, walking penalties, or intermodal accessibility costs.
\end{itemize}

Thus, the effective accessibility cost between an origin \(o\) and a destination \(u\) is:

\begin{equation}
F(o,u)=\min_{p \in P_{ou}} \sum_{(i,j)\in p} f_{ij}
\end{equation}

where \(P_{ou}\) is the set of feasible multimodal paths connecting \(o\) and \(u\).

Under this formulation, accessibility is interpreted not simply as geographic proximity, but as the capacity to overcome accumulated spatial friction within the coupled urban mobility system.

Consequently, accessibility in the model emerges from the interaction between spatial proximity and operational transport friction.

\subsubsection{Intralayer and Interlayer Friction}

Spatial friction was differentiated between intralayer movement and interlayer transfers.

For movements occurring within the same transport layer \(\alpha\), intralayer friction is defined as:

\begin{equation}
f^{[\alpha]}_{ij}=t^{[\alpha]}_{ij}+\psi^{[\alpha]}_{ij}
\end{equation}

where:

\begin{itemize}
    \item \(t^{[\alpha]}_{ij}\) represents the travel time within layer \(\alpha\),
    \item \(\psi^{[\alpha]}_{ij}\) represents additional intralayer impedance such as walking difficulty, route discontinuities, operational variability, or localized accessibility penalties.
\end{itemize}

For transfers between different layers \(\alpha\) and \(\beta\), interlayer friction is defined as:

\begin{equation}
f^{[\alpha\beta]}_{ij}=t^{transfer}_{ij}+\psi^{[\alpha\beta]}_{ij}
\end{equation}

where:

\begin{itemize}
    \item \(t^{transfer}_{ij}\) is the transfer time between layers,
    \item \(\psi^{[\alpha\beta]}_{ij}\) represents intermodal penalties associated with modal changes, transfer accessibility, transfer discontinuities, or transfer inefficiencies.
\end{itemize}

Thus, total accessibility cost emerges from the accumulation of both intralayer and interlayer friction throughout the multilayer mobility system.

This distinction is particularly relevant for evaluating whether socially distributed infrastructures such as the UTOPÍAS produce effective functional integration or remain partially disconnected due to accumulated multimodal impedance.

\subsection{Multilayer Network Formulation}

The urban mobility system was represented as a multilayer graph:

\begin{equation}
G=(V,E,L)
\end{equation}

where:

\begin{itemize}
    \item \(V\) is the set of nodes,
    \item \(E\) is the set of edges,
    \item \(L\) is the set of layers.
\end{itemize}

The layers considered in the analysis include:

\begin{itemize}
    \item walking,
    \item Metro,
    \item Metrobús,
    \item RTP,
    \item Cablebús,
    \item UTOPÍAS.
\end{itemize}

Each node belongs to a specific layer:

\begin{equation}
v_i \in V^{[\alpha]}, \qquad \alpha \in L
\end{equation}

The edge set is composed of intralayer and interlayer edges:

\begin{equation}
E = E^{intra} \cup E^{inter}
\end{equation}

where intralayer edges are defined as:

\begin{equation}
E^{intra}=\{(i,j): i,j \in V^{[\alpha]}\}
\end{equation}

and interlayer edges are defined as:

\begin{equation}
E^{inter}=\{(i,j): i \in V^{[\alpha]}, j \in V^{[\beta]}, \alpha \neq \beta\}
\end{equation}

representing transfers between transport modes and walking-access connections.

The weight associated with each edge is:

\begin{equation}
w_{ij}=
\begin{cases}
t^{walk}_{ij}, & \text{if } (i,j) \in E^{walk} \\
t^{pt}_{ij}, & \text{if } (i,j) \in E^{pt} \\
t^{transfer}_{ij}, & \text{if } (i,j) \in E^{inter}
\end{cases}
\end{equation}

Shortest-path travel times were computed using Dijkstra’s algorithm:

\begin{equation}
D(o,u)=\min_{p \in P_{ou}} \sum_{(i,j)\in p} w_{ij}
\end{equation}

where \(D(o,u)\) represents the minimum travel-time cost between an origin node \(o\) and a destination UTOPÍA \(u\).

\subsection{Connectivity and Reachability}

Global multilayer connectivity was evaluated through the relative size of the largest connected component:

\begin{equation}
C_{max}=\frac{|V_{LCC}|}{|V|}
\end{equation}

where:

\begin{itemize}
    \item \(|V_{LCC}|\) is the number of nodes in the largest connected component,
    \item \(|V|\) is the total number of nodes in the multilayer graph.
\end{itemize}

The set of reachable UTOPÍAS from an origin node \(o\) under the travel-time threshold \(\tau\) is:

\begin{equation}
R(o,\tau)=\{u \in U: D(o,u)\leq \tau\}
\end{equation}

and the number of reachable UTOPÍAS is defined as:

\begin{equation}
N_R(o,\tau)=|R(o,\tau)|
\end{equation}

This formulation allows the distinction between:

\begin{itemize}
    \item the structural presence of infrastructure,
    \item potential accessibility,
    \item and effective functional integration within the urban mobility system.
\end{itemize}

In this sense, certain UTOPÍAS may remain geographically proximate yet functionally distant due to the accumulation of operational and intermodal spatial friction within the coupled urban mobility system.

\subsection{Model Limitations}

The spatial friction model implemented in this study is based on static GTFS operational schedules and therefore represents a temporally aggregated approximation of multimodal accessibility conditions.

Although GTFS-derived travel times provide a robust operational representation of the public transport system, the model does not dynamically account for temporal variability throughout the day. Consequently, several important factors affecting effective accessibility are not explicitly incorporated into the current framework, including:

\begin{itemize}
    \item peak-hour congestion,
    \item vehicle overcrowding,
    \item demand-induced service saturation,
    \item waiting-time variability,
    \item service irregularities,
    \item operational disruptions,
    \item or temporal fluctuations in transfer efficiency.
\end{itemize}

Under this formulation, accessibility and spatial friction are evaluated as static structural properties of the coupled urban mobility system rather than as fully time-dependent dynamic processes.

This distinction is particularly important in the context of Mexico City's highly heterogeneous mobility conditions, where effective travel times and operational accessibility may vary substantially across different temporal windows of the day.

Future developments of the framework could incorporate dynamic or time-dependent multilayer accessibility models using real-time operational data, service frequency distributions, passenger demand, or temporal congestion profiles in order to capture the evolving character of urban mobility friction and functional accessibility.

\section{Pre-existing Centrality and Population Coverage Indicators}

To evaluate the territorial context in which UTOPÍAS are embedded, a set of indicators was developed to characterize both the pre-existing urban structure and the population reach of social infrastructure. Conceptually, the Pre-existing Centrality Index (PCI) and its neighborhood counterpart (WPCI) capture dimensions of structural centrality derived from economic activity, population concentration, and transport accessibility. In contrast, the Integrated Population Coverage Index (CI) and its neighborhood equivalent (WCI) quantify the territorial penetration of UTOPÍAS through isochrone-based measures of population coverage. This distinction is analytically relevant because it allows the assessment of whether UTOPÍAs are located within pre-existing urban centralities or whether they extend their influence into areas that were not previously characterized by high levels of centrality. By separating structural centrality from population coverage, the proposed framework provides a basis for examining the relationship between social infrastructure, accessibility, and the potential emergence of new functional centers within the urban socioecological system.

To evaluate the spatial context in which UTOPÍAS are embedded, a set of indicators describing pre-existing centrality, neighborhood centrality, and population coverage were constructed at the hexagonal level.

\subsection{Transport Accessibility}

A transport accessibility indicator was defined as the total number of transport service points available within each hexagon:

\begin{equation}
T_i=
TP^{Informal}_i+
TP^{Grande}_i+
TP^{Chico}_i+
Cablebus_i
\end{equation}

where:

\begin{itemize}
\item $TP^{Informal}_i$ is the number of informal transport stops;
\item $TP^{Grande}_i$ is the number of large-capacity public transport stops;
\item $TP^{Chico}_i$ is the number of small-capacity public transport stops;
\item $Cablebus_i$ is the number of Cablebús stations.
\end{itemize}

\subsection{Variable Normalization}

To allow the aggregation of variables measured on different scales, a Min-Max normalization procedure was applied:

\begin{equation}
X_i^{*}
=
\frac{
X_i-\min(X)
}{
\max(X)-\min(X)
}
\end{equation}

where $X_i^{*}$ denotes the normalized value of variable $X$ for hexagon $i$.

The following variables were normalized:

\begin{equation}
A_i^{*}
\end{equation}

(Economic Abundance)

\begin{equation}
D_i^{*}
\end{equation}

(Economic Diversity)

\begin{equation}
P_i^{*}
\end{equation}

(Total Population)

\begin{equation}
T_i^{*}
\end{equation}

(Transport Accessibility)

\subsection{Pre-existing Centrality Index (PCI)}

The Pre-existing Centrality Index (PCI) was defined as the arithmetic mean of the four normalized dimensions:

\begin{equation}
PCI_i
=
\frac{
A_i^{*}
+
D_i^{*}
+
P_i^{*}
+
T_i^{*}
}{4}
\end{equation}

where:

\begin{itemize}
\item $A_i^{*}$ represents normalized economic abundance;
\item $D_i^{*}$ represents normalized economic diversity;
\item $P_i^{*}$ represents normalized population;
\item $T_i^{*}$ represents normalized transport accessibility.
\end{itemize}

Consequently,

\begin{equation}
0 \leq PCI_i \leq 1
\end{equation}

with higher values indicating greater pre-existing functional centrality.

\subsection{Neighborhood Pre-existing Centrality Index (WPCI)}

To account for the spatial context surrounding each hexagon, a neighborhood centrality indicator was computed using the five nearest neighboring hexagons obtained through a k-nearest-neighbors (kNN) procedure.

Let:

\begin{equation}
N_i=
\{j_1,j_2,\ldots,j_k\}
\end{equation}

be the set of neighboring hexagons associated with hexagon $i$, where:

\begin{equation}
k=12
\end{equation}

The Neighborhood Pre-existing Centrality Index (WPCI) is defined as:

\begin{equation}
wPCI_i =
\frac{
\sum_{j \in N_{12}(i)} PCI_j\,d_{ij}^{-1}
}{
\sum_{j \in N_{12}(i)} d_{ij}^{-1}
}
\end{equation}

The spatially weighted PCI (wPCI) was calculated using the twelve nearest neighboring hexagons (k = 12), excluding the focal hexagon. Neighboring PCI values were weighted by the inverse of their distance $$(p = 1)$$, assigning greater influence to spatially closer hexagons. Here, $d_ij$ denotes the distance between the focal hexagon $i$ and neighboring hexagon $j$.

\subsection{Population Coverage Indicators}

Population coverage associated with UTOPÍAS was evaluated using isochrone-based population estimates.

Coverage within a 5-minute threshold was defined as:

\begin{equation}
CR5_i
=
\frac{
Iso5_i
}{
P_i
}
\end{equation}

where:

\begin{itemize}
\item $Iso5_i$ is the population contained within the 5-minute isochrone;
\item $P_i$ is the total population of hexagon $i$.
\end{itemize}

Similarly, coverage within a 10-minute threshold was defined as:

\begin{equation}
CR10_i
=
\frac{
Iso10_i
}{
P_i
}
\end{equation}

where $Iso10_i$ represents the population contained within the 10-minute isochrone.

Coverage values were constrained to the interval:

\begin{equation}
0 \leq CR5_i \leq 1
\end{equation}

\begin{equation}
0 \leq CR10_i \leq 1
\end{equation}

\subsection{Integrated Population Coverage Index}

To capture both immediate and extended service areas, an Integrated Population Coverage Index was constructed.

A greater weight was assigned to the 5-minute isochrone because it represents stronger spatial proximity and greater likelihood of use.

\begin{equation}
CI_i
=
0.6 \times CR5_i
+
0.4 \times CR10_i
\end{equation}

where:

\begin{itemize}
\item $CR5_i$ represents immediate population coverage;
\item $CR10_i$ represents extended population coverage.
\end{itemize}

\subsection{Neighborhood Integrated Coverage}

An analogous neighborhood indicator was computed as:

\begin{equation}
WCI_i
=
\frac{1}{k}
\sum_{j \in N_i}
CI_j
\end{equation}

where $WCI_i$ represents the average integrated coverage observed among the neighboring hexagons.

\subsection{Percentile-Based Ranking}

To facilitate interpretation, percentile ranks were computed for each indicator.

For a given variable $X$:

\begin{equation}
Pct(X_i)
=
\frac{
Rank(X_i)
}{
n
}
\end{equation}

where:

\begin{itemize}
\item $Rank(X_i)$ is the rank position of observation $i$;
\item $n$ is the total number of hexagons.
\end{itemize}

Thus,

\begin{equation}
0 \leq Pct(X_i) \leq 1
\end{equation}

where values approaching 1 indicate that a hexagon outperforms a larger proportion of the urban system.

\subsection{Description of Centrality and Coverage Indicators}

The following indicators were computed to characterize the territorial context, neighborhood structure, and population reach associated with each hexagonal unit within the study area.

\begin{itemize}

\item \textbf{Pre-existing Centrality Index (PCI).}

The PCI is a composite indicator that measures the degree of pre-existing functional centrality of a hexagon. It is calculated as the arithmetic mean of four normalized dimensions: economic abundance, economic diversity, population, and transport accessibility. Higher PCI values indicate locations that concentrate economic activity, population, and transportation infrastructure prior to the implementation of UTOPÍAs.

\item \textbf{PCI Percentile Rank ($PCI_{pct}$).}

The PCI percentile rank expresses the relative position of a hexagon within the distribution of PCI values across the entire study area. Values range from 0 to 1, where higher values indicate that a hexagon exhibits greater pre-existing centrality than a larger proportion of the urban system.

\item \textbf{Neighborhood Pre-existing Centrality Index ($WPCI$).}

The WPCI measures the average level of pre-existing centrality in the immediate spatial surroundings of a hexagon. It is calculated as the mean PCI value of the five nearest neighboring hexagons and represents the local territorial context in which a hexagon is embedded.

\item \textbf{Distance-weighted Neighborhood Pre-existing Centrality Index ($WPCI_{weighted}$).}

The weighted WPCI extends the neighborhood centrality concept by incorporating the influence of distance. It is computed using the twelve nearest neighboring hexagons, with inverse-distance weights assigned such that closer neighbors exert greater influence than more distant ones. This indicator captures the broader territorial structure surrounding each location while preserving spatial proximity effects.

\item \textbf{Weighted Neighborhood Centrality Percentile Rank ($WPCI_{weighted,pct}$).}

The percentile rank of the weighted neighborhood centrality index expresses the relative position of a hexagon within the distribution of distance-weighted neighborhood centrality values. Higher values indicate that a location is embedded within one of the most structurally central territorial environments of the study area.

\item \textbf{Five-minute Population Coverage Rate ($CoverageRate_{5}$).}

This indicator measures the proportion of the total population located within the five-minute isochrone associated with a UTOPÍA. It represents immediate territorial reach and captures the population that can potentially access the infrastructure within a short travel-time threshold.

\item \textbf{Ten-minute Population Coverage Rate ($CoverageRate_{10}$).}

This indicator measures the proportion of the total population located within the ten-minute isochrone associated with a UTOPÍA. Compared to the five-minute coverage rate, it reflects a broader area of influence and captures the extended service reach of the infrastructure.

\item \textbf{Integrated Population Coverage Index ($CoverageIntegrated$).}

The integrated coverage index combines immediate and extended population reach into a single measure. It is computed as a weighted average of the five-minute and ten-minute coverage rates, assigning greater importance to the five-minute isochrone due to its stronger representation of spatial proximity and accessibility. Higher values indicate greater territorial penetration of social infrastructure across the surrounding population.

\end{itemize}

\end{document}